\documentclass[hidelinks]{article}

\usepackage{arxiv}

\usepackage[utf8]{inputenc} % allow utf-8 input
\usepackage[T1]{fontenc}    % use 8-bit T1 fonts
\usepackage{hyperref}       % hyperlinks
\usepackage{url}            % simple URL typesetting
\usepackage{booktabs}       % professional-quality tables
\usepackage{amsfonts}       % blackboard math symbols
\usepackage{nicefrac}       % compact symbols for 1/2, etc.
\usepackage{microtype}      % microtypography
\usepackage{graphicx}
\usepackage{natbib}
\usepackage{doi}

\usepackage{amsmath,amssymb}

\usepackage{algorithmic}
\usepackage[linesnumbered,ruled]{algorithm2e}
\usepackage{listings}
\usepackage{subcaption}
\usepackage{textcomp}

\usepackage{verbatim}
\usepackage{appendix}
\usepackage{placeins}

\usepackage[dvipsnames,svgnames,x11names]{xcolor}

\definecolor{terminalbackground}{rgb}{0.95,0.95,0.95}

\definecolor{codebackground}{rgb}{1,1,1}      % White background color
\definecolor{codecomments}{rgb}{0.3,0.3,0.3}  % Dark gray color for comments
\definecolor{codestrings}{rgb}{0,0.5,0}       % Dark green color for strings
\definecolor{codekeywords}{rgb}{0,0,0.8}      % Dark blue color for keywords

\lstdefinestyle{mystyle}{
    backgroundcolor=\color{codebackground},   % Set white background color
    basicstyle=\small\color{black}\ttfamily,  % Set smaller font size and black text
    commentstyle=\color{codecomments},        % Set comment style
    keywordstyle=\color{codekeywords},        % Set keyword style
    stringstyle=\color{codestrings},          % Set string style
    numbers=left,                             % Show line numbers on the left
    numberstyle=\small\color{gray},           % Set number style
    numbersep=5pt,                            % Set distance from numbers to code
    breaklines=true,                          % Enable line breaking
    breakatwhitespace=true,                   % Break only at whitespace
    showstringspaces=false,                   % Do not show spaces in strings
    tabsize=4,                                % Set tab size
    language=Python,                          % Set default language
    captionpos=b,
}

\lstdefinestyle{terminal}{
    backgroundcolor=\color{terminalbackground},
    basicstyle=\ttfamily\small,
    breaklines=true,
    breakatwhitespace=true,
    frame=single,
    framesep=3pt,
    framerule=0pt,
    keywordstyle=\color{blue},
    commentstyle=\color{codegray},
    numbersep=5pt,
    showspaces=false,
    showstringspaces=false,
    showtabs=false,
    tabsize=4,
    captionpos=b,
}

\title{Multiway Alignment of Political Attitudes}

\author{ \href{https://orcid.org/0000-0002-0635-9933}{\includegraphics[scale=0.06]{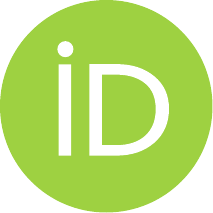}\hspace{1mm}Letizia Iannucci} \\
	Department of Computer Science\\
	Aalto University\\
	\And
	\href{https://orcid.org/0000-0001-9776-731X}{\includegraphics[scale=0.06]{orcid.pdf}\hspace{1mm}Ali Faqeeh} \\
	Department of Computer Science\\
	Aalto University\\
    \And
	\href{https://orcid.org/0000-0002-2381-6876}{\includegraphics[scale=0.06]{orcid.pdf}\hspace{1mm}Ali Salloum} \\
	Department of Computer Science\\
	Aalto University\\
    \And
	\href{https://orcid.org/0000-0002-3279-8710}{\includegraphics[scale=0.06]{orcid.pdf}\hspace{1mm}Ted Hsuan Yun Chen} \\
	Department of Environmental Science and Policy\\
	George Mason University\\
    \And
	\href{https://orcid.org/0000-0003-2049-1954}{\includegraphics[scale=0.06]{orcid.pdf}\hspace{1mm}Mikko Kivelä} \\
	Department of Computer Science\\
	Aalto University\\
}

\date{}

\renewcommand{\headeright}{}
\renewcommand{\undertitle}{}
\renewcommand{\shorttitle}{Multiway Alignment of Political Attitudes}

\hypersetup{
pdftitle={Multiway Alignment of Political Attitudes},
pdfsubject={},
pdfauthor={Letizia Iannucci, Ali Faqeeh, Ali Salloum, Ted Hsuan Yun Chen, Mikko Kivelä},
pdfkeywords={},
}

\begin{document}
\maketitle

\begin{abstract}
    The related concepts of partisan belief systems, issue alignment, and partisan sorting are central to our understanding of politics. These phenomena have been studied using measures of alignment between pairs of topics, or how much individuals' attitudes toward a topic reveal about their attitudes toward another topic. We introduce a higher-order measure that extends the assessment of alignment beyond pairs of topics by quantifying the amount of information individuals' opinions on one topic reveal about a set of topics simultaneously. Applying this approach to legislative voting behavior shows that parliamentary systems typically exhibit similar multiway alignment characteristics, but can change in response to shifting intergroup dynamics. In American National Election Studies surveys, our approach reveals a growing significance of party identification together with a consistent rise in multiway alignment over time.
\end{abstract}

\section{Introduction}

The concept of belief systems, which encompass configurations of ideas and attitudes constrained by some form of interdependence~\citep{converse2006nature}, has long been central to the study of politics. Extensive evidence suggests that the emergence of political belief systems influences political affiliations, attitudes, judgments, behaviors, and the perception of facts~\citep{converse2006nature, van2018partisan}. Strong political belief systems may contribute to the formation of entrenched partisan identities, which threatens the democratic process, leading to conflicts and hindering the ability of individuals to understand outgroup views~\citep{hetherington2009polarization}.

Constraint between attitudes is often understood as the extent to which an individual's position toward one issue predicts their position toward another~\citep{converse2006nature}. Similarly, partisan sorting  manifests as individuals taking on the issue positions of the parties they identify with \citep{baldassarri2008partisans}. The emergence of partisan belief systems, then, can be investigated by considering how attitudes toward issues are aligned, through some form of statistical association, with attitudes toward other issues (i.e., issue alignment) or toward parties (i.e., partisan alignment or issue partisanship). Previous research on partisan sorting, issue alignment, and political polarization has done so using pairwise metrics, such as correlation~\citep{baldassarri2008partisans, kozlowski2021issuecorrelation, bougher2017correlates, hetherington2009polarization,luders2024attitude} and mutual information~\citep{chen2021polarization}.

However, we have no reason to believe constraints only exist at the pairwise level -- a belief \textit{system} contains a collection of issue pairs, which are unlikely to be independent of one another. While existing pairwise approaches intuitively capture the concept of constraints in belief systems, and can even be used to study collections of issue pairs \citep[e.g.,][]{luders2024attitude,fishman2022change,chen2021polarization}, they are unable to describe alignment jointly across multiple topics (i.e., issues and parties), missing the potential for there to be multidimensional constraints. Exclusively focusing on pairwise associations may fail to capture the richer structure of higher-order interdependencies that exists among sets of political topics -- such as how voters align themselves on multiple issues simultaneously or how legislators position themselves in a multidimensional policy space -- which, as we show below, is not simply the aggregation of pairwise constraints.

To better understand how belief systems shape political attitudes, we propose the careful consideration of how topics can be \textit{jointly aligned}, and we develop a measure of multiway alignment to study this phenomenon. By pursuing multiway alignment, we are emphasizing that real-world systems extend beyond pairwise interactions~\citep{battiston2020networks, bougher2017correlates, levin2021dynamics}, a notion that intuitively flows from our existing understanding of politics. An example of this can be found in the liberal versus conservative divide in U.S. politics~\citep{bougher2017correlates}. Numerous studies on partisan sorting demonstrate that individuals who identify as liberal or conservative are more likely to align themselves with the party that offers policy bundles consistent with their ideological leanings~\citep{bafumi2009new, baldassarri2008partisans, bougher2017correlates}. More recently, what had primarily been a process of partisan alignment has also extended to entrenched ideological or issue alignment \citep{kozlowski2021issuecorrelation}. This phenomenon is not limited to the U.S., as similar trends towards polarization have also been observed in European countries~\citep{westwood2018tie, flores2022politicians}. These observations suggest that the study of polarization and partisan identities should focus on cleavages arising from the alignment of multiple issues.

A common way to measure the degree of agreement among multiple individuals and across multiple issues is Kendall's W~\citep{gibbons1990rank}, which quantifies the agreement across different rankings of the same set of objects. Kendall's W, however, is a linear function of the averages of the pairwise Spearman's rank correlation coefficients between every possible pair~\citep{gibbons1990rank}. As we show in the following section, the average of pairwise scores does not appropriately describe systems at higher orders.
Alternatively, when examining ideological cleavages across multiple topics and multidimensional constraints, researchers can use multivariate measures that capture global characteristics of joint distributions \citep[e.g.,][]{rosas2019oinfo}, such as the extent of information shared among variables and the level of interdependence, through the entropies of the stochastic variables. We document these in Appendix~\ref{a:other-measures}.
Some of these measures, however, present applicational challenges in that they do not have directly correspondent bivariate analyses. Moreover, the use of unnormalized measures complicates the interpretability of the scores and comparisons across different datasets.

We derive a measure of multiway alignment based on entropy and mutual information to capture higher-order interdependencies. Our proposed quantification of multiway alignment generalizes the method in~\citet{chen2021polarization}, allowing to quantify alignment across any number of topics. While pairwise alignment in~\citet{chen2021polarization} is quantified by the amount of information that knowing someone's opinion on one topic provides on their opinion on another topic, we measure the amount of information that \textit{knowing someone's opinion on one topic provides on their opinion on the remaining $n-1$ topics}. Additionally, we ensure the significance of our proposed score by comparing it with the amount of alignment that can be attributed to chance. 

Our approach to multiway alignment analysis facilitates understanding the interplay between various divides and beliefs. Our analysis of data from the \citet[ANES;][]{anes} reveals the increasingly substantial influence of party identification in shaping opinion dynamics among the U.S. public. To facilitate further exploration of our findings and application to other systems, we provide the Python package \texttt{multiway\_alignment} (see Appendix~\ref{a:pip-package}), 
which lets researchers apply our multiway alignment methodology. 

The paper proceeds as follows. 
First, we establish the theoretical framework that extends the concept of alignment beyond the pairwise to effectively model higher-order interdependencies. Then, we detail our approach to quantifying multiway alignment. We proceed to demonstrate the broad applicability of our measure to a variety of social and political systems and discuss general patterns of higher-order alignment that may be observed in real-world systems. We then discuss our findings from the analysis of data from the \citet[ANES;][]{anes}. Finally, the last section summarizes our contributions and outlines future research directions.

\section{Importance of Multiway Alignment}
\label{s:theoretical_framework}

We propose a multiway alignment measure that gauges the extent to which individuals’ attitudes toward different topics collapse into a single divide. This concept goes beyond examining alignment over pairs of topics and instead investigates the alignment of opinions over an array of topics and the extent to which consensus among individuals transcends the specific subjects being discussed. To do so, we define the idea of \textit{consensus partition} to capture the multidimensional nature of social systems. Specifically, a consensus partition refers to a division of individuals into groups (i.e., a consensus group) based on how they share opinions across multiple topics. As such, these partitions comprise groups of individuals who hold the same constellation of positions across a set of topics. We further formalize this concept in Section~\ref{s:methods}.

\begin{figure}[t]
    \centering
    \includegraphics[width = \textwidth]{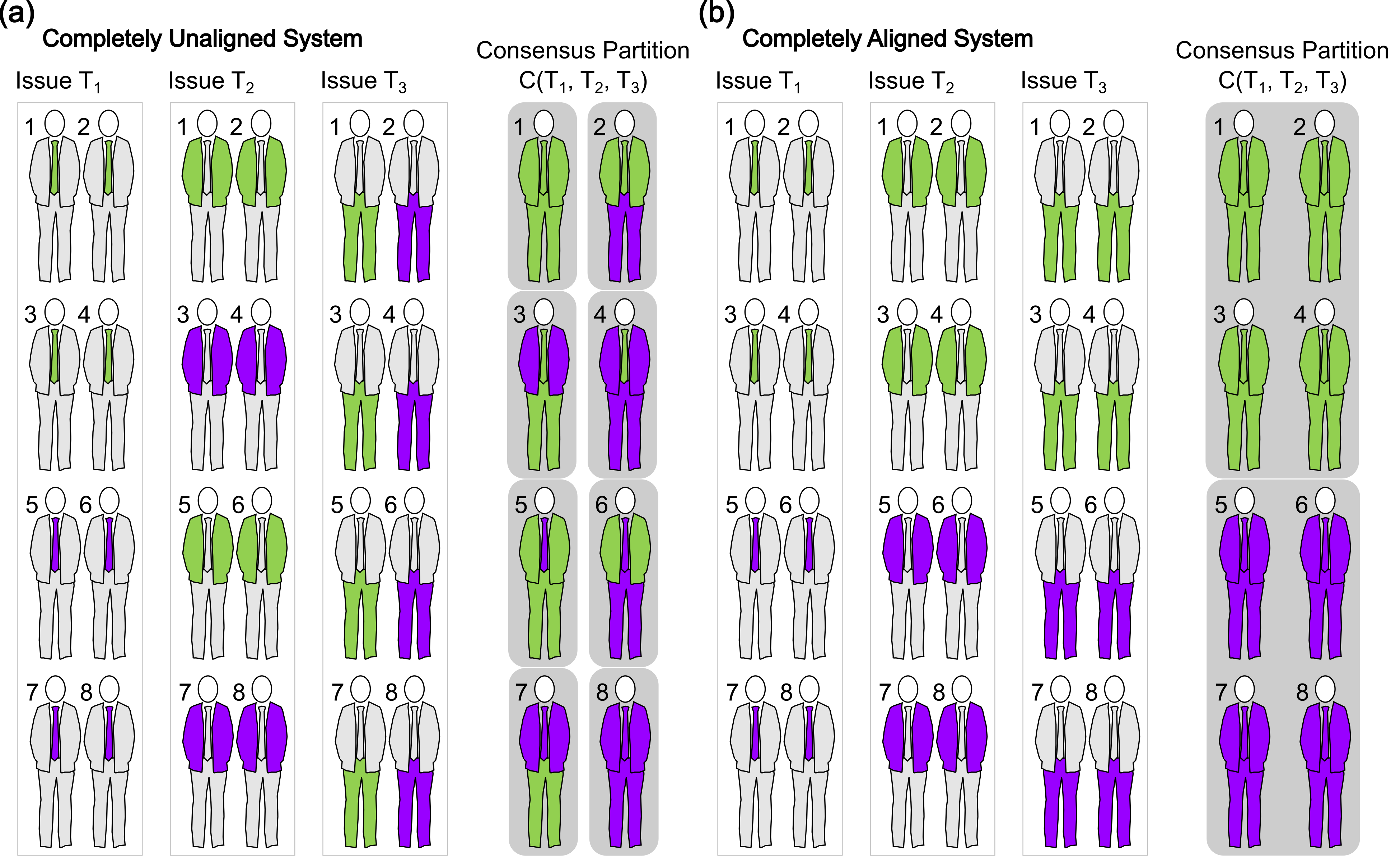}
    \caption{Illustration of consensus partitions based on individuals' preferences toward three issues (tie color $T_1$, shirt color $T_2$, and pants color $T_3$). Panel \textbf{(a)} illustrates a system with no alignment, where the opinion groups found within each issue become fragmented into smaller consensus groups as additional issues are considered. Panel \textbf{(b)} illustrates a system with perfectly aligned positions on the three issues. In this case, the resulting consensus partition exactly matches the opinion groups for each issue.
    }
    \label{fig:consensus-partition}
\end{figure}

Consensus partitions matter to our understanding of alignment across multiple topics in that they capture the extent to which ideological or partisan cleavages sort individuals into groups that have overlapping preferences, which depends on the extent to which these cleavages are either aligned or cross-cutting. Specifically, when individuals can take one of two positions on any political topic, a system with $n$ topics means there are $2^n$ possible constellations of positions (i.e., consensus groups). When cleavages are highly aligned, individuals will sort into a consensus partition such that only few of these consensus groups actually meaningfully exist (i.e., most groups are very small or nonexistent). At the extreme where cleavages are perfectly aligned, there will only be two consensus groups that have no shared attitudes between them. On the other hand, when cleavages are cross cutting, there will be many 
relatively balanced consensus groups in society, and even relatively like-minded individuals will rarely agree on everything, and in turn will not disagree with their outgroup on everything. In this respect, cross-cutting cleavages create a more complex identity space, as the diversity of stances results in a tapestry of identities that defy simple categorization (e.g., someone might be economically conservative but socially liberal, or support environmental policies while holding different views on immigration).

We illustrate this concept more explicitly in Figure~\ref{fig:consensus-partition} with two extreme cases involving eight individuals, and three issues with binary choices (i.e., green or purple ties, shirts, and pants). The absence of alignment between the three issues (Figure~\ref{fig:consensus-partition}a) yields the least aligned consensus partition of maximally balanced consensus groups ($2^3 = 8$ groups with one person in each group). In this case, even when the issues themselves are divisive, the impact of exclusive identities is mitigated since opinion groups toward one issue are not related to opinion groups toward the other issues. Individuals can find common ground despite disagreement on specific issues -- for all individuals in Figure~\ref{fig:consensus-partition}a, there is only one person with whom they have no shared preference at all.

On the other extreme, Figure~\ref{fig:consensus-partition}b shows an instance of consistent alignment of opinion groups across the three issues, indicating high interdependence among the issues. This means the individuals in the two different consensus groups hold conflicting views on all three issues. Their consistent opposition implies that the two groups' beliefs are structured using a similar rationale, albeit applied in opposite directions. The implication here is that, despite the even split of individuals within each issue in both examples, the perfect alignment of issues in the latter case yields large and entrenched consensus groups of individuals who always agree with their ingroup and never with the outgroup.

In general, each political topic either introduces potential for cross-cutting cleavages or confirms (i.e., aligns with) the existing consensus partition. To illustrate this, consider the addition of a fourth issue, e.g., preference toward shoes, which again has four green positions and four purple positions. If shoe preferences align with existing preferences for pants, shirts, and ties, all individuals who prefer green pants, green shirts, and green ties will also agree on the same color of shoes. This alignment would strengthen cohesive preferences across the wider array of issues without altering the existing consensus groups -- there would still be two groups even though there are $2^4 = 16$ possible ones. Conversely, if shoe preferences do not align with other issues under consideration, individuals who prefer green pants, green shirts, and green ties may differ in their opinions on shoe color. Some may opt for green shoes while others choose purple shoes. This lack of alignment would fragment the previously established consensus partition, resulting in additional, balanced consensus groups.

\subsection{Pairwise Alignment Misses Higher-Order Patterns}

As noted above, while one could use pairwise alignment measures between all pairs of topics to estimate the amount to which all of them are aligned -- for the example in Figure~\ref{fig:consensus-partition}, this would amount to computing three values, one between each of the three pairs -- it would be insufficient for inferring the existence of higher-order constraints. Measures such as correlation or mutual information can describe the extent of pairwise alignment, but they ignore the potential dependencies that may exist among three or more topics simultaneously. Therefore, summarizing a set of pairwise scores, for example by summing or averaging, to describe a higher-order constraint can lead to an inaccurate representation of the higher-order constraints.

\begin{figure}[t]
    \centering
    \includegraphics[width = 0.5\textwidth]{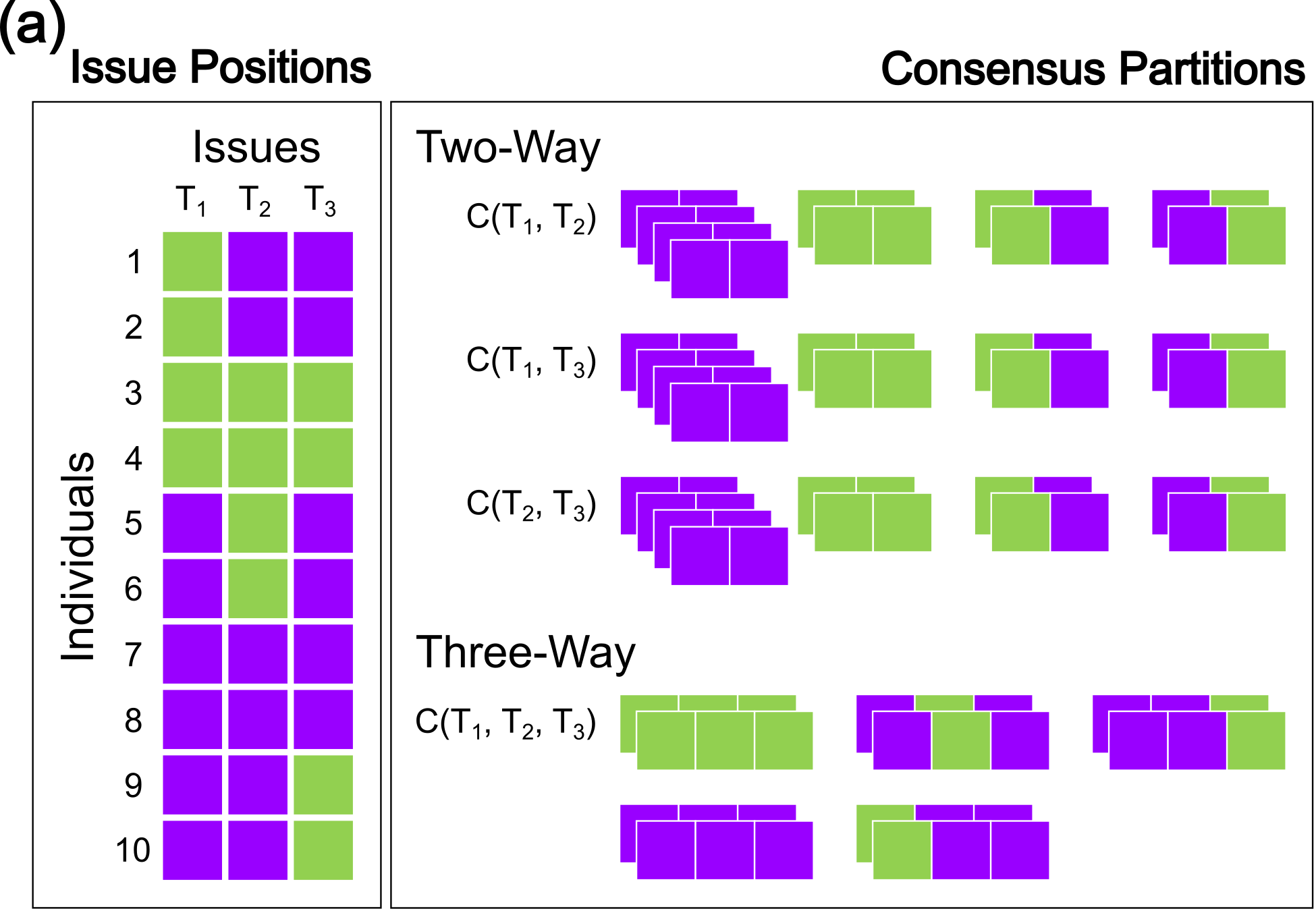}%
    \includegraphics[width = 0.5\textwidth]{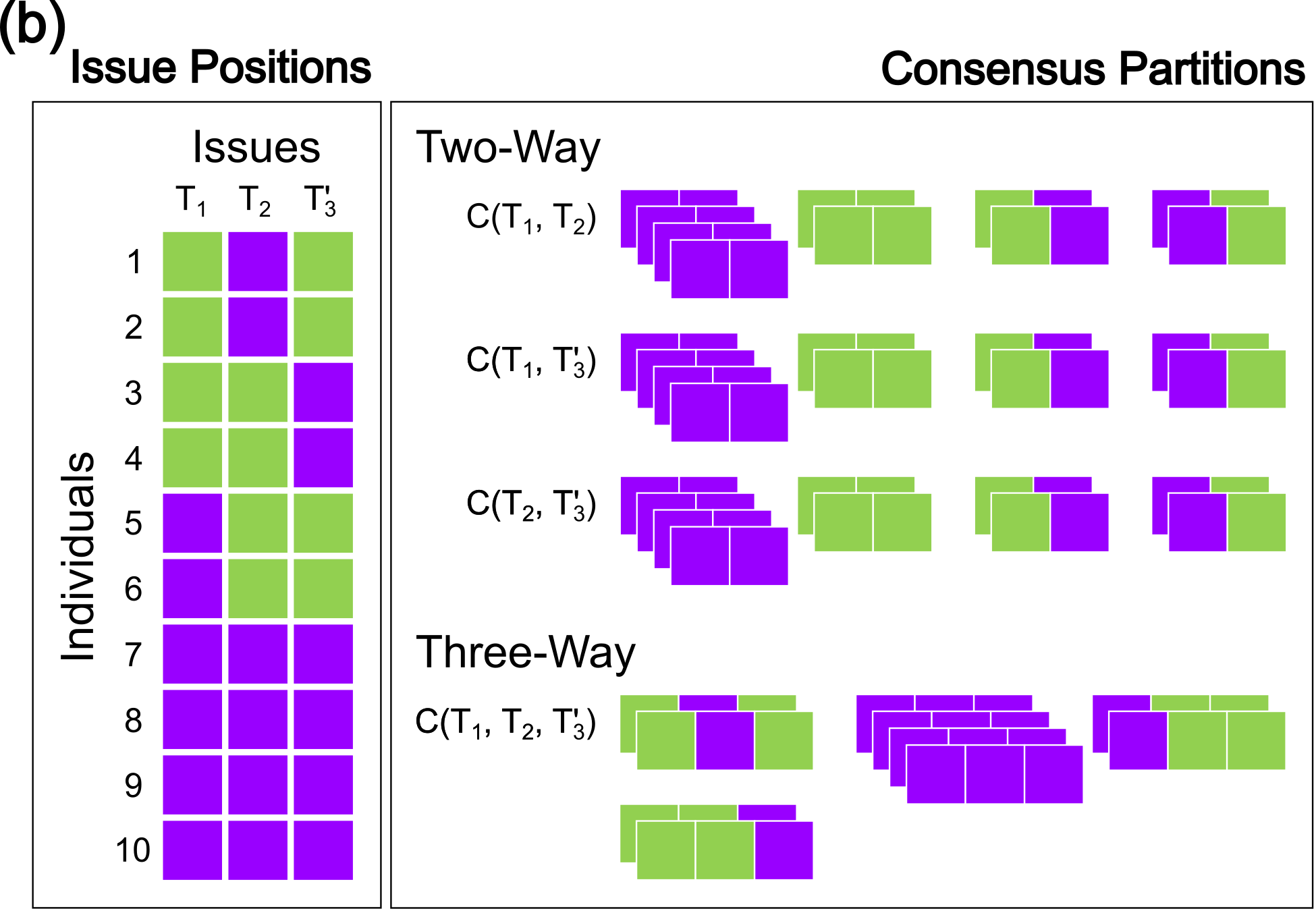}
    \caption{
    An example where 2-way alignment measures cannot capture multiway alignment.
    Panel \textbf{(a)} and Panel \textbf{(b)} show two systems of ten individuals exhibiting the exact same set of pairwise associations but different multiway alignment over three topical issues ($T_1, T_2, T_3$). The squared cells represent individuals' preference on each issue ("agree" in purple, and "disagree" in green). Issue $T_1$ and issue $T_2$ are the same in Panel \textbf{(a)} and in Panel \textbf{(b)}, whereas $T_3$ and $T'_3$ generate different opinion partitions. The consensus partitions $C(T_2, T_3)$ and $C(T_2, T'_3)$ show that the amount of pairwise alignment between issue $T_2$ and $T_3$ and between $T_2$ and $T'_3$ is exactly the same, since both pairs generate the same number of consensus groups, and individuals are partitioned with the same relative frequencies. The same holds for $T_1$ and $T_3$ in Panel \textbf{(a)} and for $T_1$ and $T'_3$ in Panel \textbf{(b)}. Yet, $C(T_1, T_2, T_3)$ in Panel \textbf{(a)} and $C(T_1, T_2, T'_3)$ in Panel \textbf{(b)} demonstrate that the pairwise alignments do not uniquely determine multiway alignment: $C(T_1, T_2, T_3)$ in Panel \textbf{(a)} and $C(T_1, T_2, T'_3)$ in Panel \textbf{(b)} describe two different amounts of 3-way alignment, lower in Panel \textbf{(a)} and higher in Panel \textbf{(b)}.
    }
    \label{fig:schem-double}
\end{figure}

We illustrate two potential shortcomings of pairwise measures using the examples in Figure~\ref{fig:schem-double}. First, the level of alignment in a system at the pairwise level does not necessarily reflect the level of alignment at the triad and higher. Consider Figure~\ref{fig:schem-double}a, which illustrates three issues ($T_1$, $T_2$, and $T_3$) with binary choices
(represented by color, i.e., green or purple). The consensus partitions involving only two issues (i.e., $C(T_1,T_2)$, $C(T_1,T_3)$, and $C(T_2,T_3)$) show that any pair of issues partitions the population into four consensus groups. Because the division of individuals into the groups (i.e., the 2-way consensus partition) is the same for any pair of issues, all pairs of issues exhibit the same amount of alignment. In other words, any alignment measure based on the resulting partition will yield the same score for all pairs of issues. However, when considering all three issues simultaneously, we find that the overall alignment of the system cannot be captured solely by examining the scores obtained at the pairwise level. Specifically, in this case, the consensus partition collectively induced by $T_1$, $T_2$, and $T_3$ does not preserve the group cleavages obtained from the $2$-combinations. This is because each pair of topics initially creates a larger consensus group of four individuals (taking the purple position for both issues), but the introduction of the third topic always ends up dividing this group into two smaller groups. Since the third topic is always cross-cutting one of the cleavages that exist at the pairwise level, the number of consensus groups increases, showing higher diversity of stances and lower 3-way alignment compared to what is observed at the pairwise level.

Second, relying solely on the assessment of pairwise alignment may hinder our ability to differentiate between different levels of higher-order alignment arising from the same pairwise relationships. We demonstrate this with Figure~\ref{fig:schem-double}b, which shows the same $T_1$ and $T_2$ as in Panel (a), but changes $T_3$ to a different $T'_3$. The systems in Panel (a) and in Panel (b) look identical at the pairwise level, as the consensus partitions $C(T_1,T_3)$ and $C(T_1,T'_3)$, and $C(T_2,T_3)$ and $C(T_2,T'_3)$ are equivalent. However, considering all three issues simultaneously yields different consensus partitions $C(T_1,T_2,T_3)$ and $C(T_1,T_2,T'_3)$. Therefore, the systems in Panel (a) and in Panel (b) exhibit different levels of 3-way alignment, with the former being less aligned (i.e., having a greater number of consensus groups).

The possibility of observing different higher-order constraints arising from the same set of pairwise alignments underscores the necessity of extending beyond pairwise analysis, as relying solely on comparisons between pairs of topics poses the risk of inaccurately portraying the higher-order alignment of a system. By studying consensus groups defined by individuals' opinions on multiple topics, we are able to define a measure of multiway alignment that captures the overall consistency of beliefs across an entire spectrum of political topics. 

\section{Measuring Multiway Alignment}
    \label{s:methods}

In this section, we formally develop our multiway alignment measure. In Section~\ref{ss:mutual_clusters}, we formalize the idea of consensus partitions and its significance in investigating the extent to which opinion groups across multiple topics are either aligned or cross-cutting. In Section~\ref{ss:multiway_alignment_score}, we present our formulation for a multiway alignment measure and discuss its desired properties.

\subsection{Consensus Partitions}
    \label{ss:mutual_clusters}

Formally, we investigate a set of individuals $I$, and how they are divided to opinion groups $g \subseteq I$ for a given topic. We assume that each individual holds exactly one opinion in a single topic, such that the opinion groups are mutually exclusive. That is, each topic $T$ can be represented as a partition of the set of all individuals $I$, which is formally defined as 
$$
T \subseteq \{g \quad | \quad g \subseteq I\}, \quad \text{ such that } \quad \emptyset \notin T, \quad g \cap g' = \emptyset \quad \forall g \neq g' \in T,  \quad \text{ and }\quad  \bigcup_{g \in T} g = I\,.
$$

To quantify alignment in the sense of Figure~\ref{fig:consensus-partition}, we identify clusters of individuals that consistently belong to the same opinion group, i.e., are like-minded, across the given $k$ topics $T_1, \ldots, T_k$. We refer to these clusters as consensus groups defined by the set $T_1, \ldots, T_k$. 
The set of consensus groups is a consensus partition as each individual again belongs to exactly one consensus group. Formally, a consensus partition $C(T_1, \ldots, T_k)$ is 
defined as the set of non-empty intersections of the partitions defined by each topic:

\begin{equation} \label{eq:1}
    C (T_1, \ldots, T_k) = \left\{ c \subseteq I \quad | \quad \exists (g_1, \ldots, g_k), \quad g_i \in T_i \quad \text{s.t.} \quad \bigcap_{i=1}^k g_i = c, \quad c \neq \emptyset \right\},
\end{equation}
where $T_i$ represents the partition of individuals defining the opinion groups for the $i$-th topic.

Besides having a clear interpretation in terms of societal dynamics, the consensus partition also effectively represents the joint distribution of opinions across the topics under consideration. Consequently, the entropy $H(T_1, \ldots, T_k) = H(C(T_1, \ldots, T_k))$ represents the amount of information needed to describe the consensus partition, which reflects the complexity of the identity space. When individuals mostly agree with their ingroup on various topics or divides, the consensus partition has low entropy: even when the set of topics is large, the amount of information required to describe the consensus partition is minimal, as there are only few entrenched consensus groups. Conversely, a consensus partition with high entropy describes a complex identity space with cross-cutting cleavages. Because of the multitude of diverse consensus groups, the identity space is more heterogeneous, and the consensus partition requires a longer description to capture all the identities. 

The emergence of belief systems, then, can be characterized by the amount of information that each individual topic contains about the consensus partition obtained from the other topics in analysis. Specifically, in a highly interdependent system, the uncertainty in the consensus partition greatly decreases when we acquire knowledge about the opinion groups on one topic. Conversely, in the absence of constraint, knowing the opinion groups on one topic provides little explanation of the configuration of the consensus partition.

\subsection{Multiway Alignment}
\label{ss:multiway_alignment_score}

Given the interpretation of the consensus partition presented in the previous Section, we establish a quantitative measure of multiway alignment across a set of $k$ topics based on how similar the opinion partition of each topic is to the consensus partition of the $k-1$ remaining topics. In particular, for a given list of topics $T_1, \ldots, T_k$, we can quantify their multiway alignment by averaging a chosen similarity measure \textit{S} between each topic and the consensus partition defined by the remaining topics:
\begin{equation} \label{eq:general}
A_{\text{S}} = \frac{1}{k} \sum_{i=1}^{k} \text{S}(T_i, C(T_1, \ldots, T_{i-1}, T_{i+1}, \ldots, T_k)).
\end{equation}
If the similarity is based on mutual information, i.e., $\text{S}=\text{MI}$, the measure describes the average amount of information knowing someone's opinion in one topic gives about their position in all of the other topics.
As shown in Appendix~\ref{a:mi_k3}, $A_{\text{MI}}$ can be interpreted as the total correlation, adjusted by subtracting contributions from lower-order alignments. 

The measure defined in Equation~\ref{eq:general} is able to describe higher-order phenomena while still having an intuitive interpretation and  correspondence with the relevant theoretical concepts of consensus and opinion groups we developed above. Even though the definition does not restrict the choice of the similarity measure $\text{S}$, we seek to choose $\text{S}$ in a way that guarantees certain desirable properties for our measure:

\begin{enumerate}
    \item When the multiway alignment is computed for $n=2$, the multiway measure should reduce to an existing pairwise measure that could be used to quantify alignment. For example, our multiway alignment measure should reduce to the pairwise alignment measure used in \citet{chen2021polarization} if we define it in terms of normalized mutual information (NMI). This ensures consistency and comparability with existing methods.
    \item The measure should be clearly normalized and not depend on factors such as the number of individuals or imbalance of the opinions. We would like the value 1 to indicate complete alignment and 0 to indicate no alignment. This ensures that the results are easily interpretable and comparable across different contexts.
\end{enumerate}

\paragraph{Property 1} As long as the chosen S is symmetric, $A_\text{S}$ reduces to the original pairwise measure $\text{S}$ for $k=2$, since the consensus partition $C(T_j)$ is defined by topic $T_j$ alone and therefore matches opinions in $T_j$ itself. By the same reasoning above, the formula in Equation~\ref{eq:general} simplifies to exactly the measure used in~\citet{chen2021polarization} for $k=2$ when S = NMI. Yet, the formulation in Equation~\ref{eq:general} allows us to quantify also alignment across multiple topics, as the mutual information is now computed between each topic and the consensus partition defined by the remaining topics in analysis. 

\paragraph{Property 2} The normalization of mutual information is achieved by scaling $\text{MI}$ by one of its upperbounds: for example, the maximum of the entropies, the geometric average of the entropies, or the arithmetic mean of the entropies~\citep{nmi}. As a result, normalized mutual information takes on values between $0$ and $1$, where $0$ indicates independence and $1$ signifies complete agreement between the two labeling assignments that are compared. The $A_{\text{NMI}}$ score naturally inherits those properties, and therefore allows for similar interpretation. NMI, however, does not correct for agreement solely due to chance and thus may yield inflated scores, particularly when comparing groups with significant size imbalances~\citep{reduced-mi, alinormalization}. This limitation also applies to the multiway alignment score $A_{\text{NMI}}$. We mitigate this issue by using the adjusted mutual information (AMI)~\citep{nmi} instead, and all our numerical results will be presented in terms of $A_{\text{AMI}}$. Specifically, 

$$\text{AMI}(U, V) = \frac{\text{MI}(U, V) - \mathbb{E}\{\text{MI}(C)|a_1, \dots, a_m\}}{ \text{max}\{ \text{MI}(U, V) \} - \mathbb{E}\{\text{MI}(C)|a_1, \dots, a_m\} }\,,
$$
where $\text{MI}(U,V)$ is the mutual information between $U$ and $V$, $\text{max}\{ \text{MI}(U, V) \}$ is one of the upperbounds of $\text{MI}$ that can be used to normalize $\text{MI}$, $a_1, \dots, a_m$ are the marginals from the contingency table $C$ between $U$ and $V$ so that the expected mutual information $\mathbb{E}\{\text{MI}(C) \mid a_1, \dots, a_m\}$ is conditioned on the overall distributions of the cluster sizes~\citep{nmi}. 

Therefore, a multiway alignment score defined as in Equation~\ref{eq:general} with $\text{S}=\text{AMI}$ satisfies the properties stated above and provides a quantitative tool for analyzing higher-order alignment. The choice of AMI also ensures robustness by accounting for cases where pairwise alignment between opinions on one issue and a consensus partition could be attributed to random chance. We further validate the statistical significance of our multiway alignment findings by introducing a null model for comparison (see Appendix~\ref{a:nullmodel}). The null model provides the expected outcome $\langle A_{\text{null}} \rangle$ when the multiway alignment is a consequence of random chance and allows us to establish confidence intervals that account for fluctuations in the value of $A$ due to the finite number of individuals in the system.

\section{Alignment Spectrum}
    \label{s:alignment-spectrum}

In this section, we introduce the concept of the \textit{alignment spectrum} and explain how it can be used to understand higher-order alignment patterns across multiple topics. The alignment spectrum emerges from the calculation of all alignment scores across all combinations of topics. Essentially, it illustrates how political attitudes relate to each other as we expand the set of topics being considered together. In fact, within a system of $k$ topics, we can calculate several alignment scores -- specifically, pairwise alignments between each pair of topics, as well as all higher-order alignment scores that consider order $3, \ldots, k$. For example, in the case of four topics, the alignment spectrum will consist of the scores derived from pairs of topics, triplets of topics, and the full set of four topics. This results in a total of seven alignment scores that reflect how individuals' opinions relate across different subsets of the four topics. By examining these scores, we can identify patterns of alignment and study how alignment evolves as we increase the number of topics. 

\begin{figure}
    \centering
    \includegraphics[width=1\linewidth]{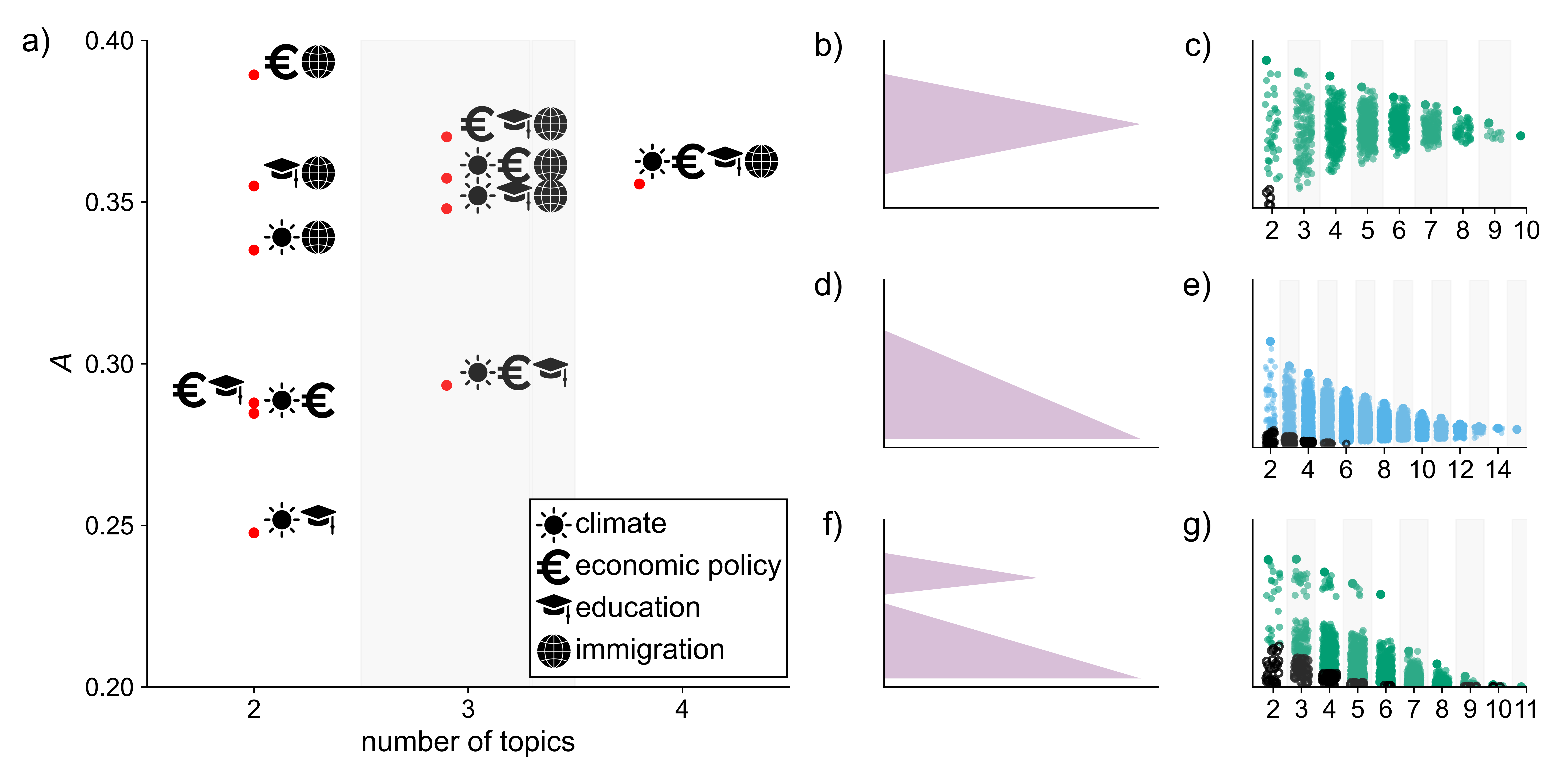}
    \caption{An example of alignment spectrum from online discussions related to four topics before Finnish elections 2023 (Panel \textbf{a}). The x-axis indicates the size of the topic combinations, which corresponds to the order of alignment, while the y-axis shows the corresponding alignment scores. Each point in the figure indicates a specific subset of topics, as indicated by the icons. Panel \textbf{b, d, f} show different fundamental patterns of higher-order alignment that we observe when analyzing real-world systems with our proposed measure. Moderate or high multiway alignment (Panel \textbf{b}) is usually observed in legislative systems, such as Finnish Parliament (Panel \textbf{c}). Panel \textbf{d} shows that, even when pairwise alignment is moderate, alignment at higher-order might be low. An example of this is shown in Panel \textbf{e}, where alignment at order $>3$ rapidly decreases towards 0. Finally, Panel \textbf{f} shows an example of mixed pattern of higher-order alignment, where some topics are aligned and others are not. This was the case in Finnish parliamentary discussions during 2020--2021 (Panel \textbf{g}). The black markers in Panel \textbf{c, e, g} show values of alignment that are not statistically significant, as they fall within the $95\%$ confidence interval obtained by the null model. In Panel \textbf{a}, all points are significant with respect to the null model.}
    \label{fig:spectrum}
\end{figure}

We show an example of an alignment spectrum in Figure~\ref{fig:spectrum}a, which illustrates the structure of online discussions around four topics (climate, economic policy, education, and immigration) in the Finnish Twittersphere leading up to the 2023 Finnish elections. The $x$-axis represents the order of alignment: 2 for pairs of topics, 3 for triplets, and 4 for the full set of topics. Each point on the plot corresponds to a specific subset of topics in the respective order, and the $y$-axis displays the alignment values for each of those subsets. The alignment spectrum summarizes different alignments in the system and shows how the patterns of alignment observed between pairs of topics extend to higher orders or weaken as more topics are considered. In the example shown in Figure~\ref{fig:spectrum}a, the level of alignment observed between opinions on economic policy and immigration remains almost the same when adding a third issue (i.e., education) and even when all four topics are included. This indicates that the alignment is robust across multiple dimensions, and individuals who have a specific stance on one topic (e.g., immigration) are likely to have predictable stances on other topics (e.g., economic policy and education). Conversely, if the alignment dropped significantly when more topics are considered, the spectrum would indicate that the pairwise alignments do not extend to a broader set of issues, revealing the existence of cross-cutting cleavages. 

In general, when using $A_{\text{AMI}}$, the theoretical upper bound of the alignment spectrum is represented by the horizontal line at $A=1$, indicating perfect alignment across all the issues. In contrast, the alignment values for a completely random system would be close to $0$ at any order. As these bounds represent two extremes of ideological coherence, the alignment spectrum of a real-world system can describe various types of constraints within this range.

Figure~\ref{fig:spectrum} illustrates two fundamental patterns that may emerge when analyzing political alignment across multiple issues using our proposed measure.
In cases of moderate or high multiway alignment, individuals tend to hold consistent views across issues, leading to a funnel-shaped alignment spectrum where multiway alignment scores remain significantly above zero even when considering all topics (Figure~\ref{fig:spectrum}b). This pattern is commonly observed in legislative systems, where the alignment across different topics reflects the dynamics of policymaking and political discourse. Figure~\ref{fig:spectrum}c shows the alignment spectrum of Finnish Parliament in 2019--2020, which demonstrates alignment that could originate from adherence to party lines, whereby members from the same party (or coalition) vote in a similar way on a wide range of issues. The same effect is also visible when analyzing voting behavior in U.S. House (see Appendix~\ref{a:results-house}).
Being a two-party system, the U.S. exhibits a much more pronounced effect of elite polarization on the spectrum of multiway alignment, and the multiway alignment score usually remains at moderate values even when considering as many as $18$ different topics. Yet, the pattern in Figure~\ref{fig:spectrum}b is not limited to legislative systems. We observe a similar alignment spectrum also for social media discussions centered around political topics before Finnish elections (Figure~\ref{fig:spectrum}a). Although the emergence of alignment within discussions around political parties is anticipated in a multi-party system, we discuss in Appendix~\ref{a:results-twitter} how the observed increase in higher-order alignment across issues signifies a shift towards a more partisan-centric political landscape.

The pattern of alignment in Figure~\ref{fig:spectrum}b occurs because individuals maintain some level of alignment across multiple topics. At the top of the spectrum, scores are high because opinion groups tend to remain similar across the issues. As the number of issues increases (from pairs to triads and beyond), the alignment score decreases. However, when opinion groups are generally aligned, this decline is gradual, forming the funnel shape. The bottom part of the funnel represents the minimum alignment observed for each number of issues. If this minimum score is significantly above zero, it suggests the existence of constraints that could span a larger set of topics. Conversely, if the system does not exhibit higher-order alignment, the alignment scores drop close to zero as more issues are included. 

This pattern of low higher-order alignment manifests as a triangular-shaped spectrum that rapidly decreases toward zero as more issues are considered (Figure~\ref{fig:spectrum}d). The lower triangle shape indicates that individuals' opinions across different topics are largely independent at higher-orders, or that the stances over larger sets of issues become cross-cutting. In this scenario, knowing someone's position on one issue might provide little to no information about their stance on a combination of other issues. Figure~\ref{fig:spectrum}e shows an example of this type of spectrum, that starts with high alignment scores for pairs of issues but rapidly declines as more issues are added.

The alignment spectrum can also exhibit a mixed pattern, where some issues are aligned while others show little to no alignment. An example of this is the distinctive double-triangle shape in Figure~\ref{fig:spectrum}f. This mixed pattern consists of an upper funnel-like section and a lower triangular section. The upper triangle, resembling a small funnel, represents the subset of issues where individuals tend to be aligned. For these issues, alignment scores remain relatively high, even as the number of issues considered increases, but only for a subset of the topics. The lower triangle mirrors a low-alignment pattern, where scores quickly approach zero as more issues are introduced. This double-triangle pattern was observed in the Finnish Parliament during the COVID-19 pandemic (Figure~\ref{fig:spectrum}g), and reveals a known phenomenon that would be missed in a purely pairwise analysis. Notably, during the COVID-19 pandemic, especially early on, Finnish parliamentary politics exhibited unconventionally high levels of cooperation between the government and opposition coalitions~\citep{lehtonen2024navigating}. This was evident in how both sides repeatedly appealed for broad cooperation, reciprocally praise each other for implemented COVID responses, and even in how the opposition withdrew their interpellation on the government's immigration policy~\citep{eduskuntapressrelease}. Despite this drastic shift from conventional politics, the changes brought by the pandemic become glaringly evident only when analyzing the patterns of alignment at higher orders, where the double-triangle shape of the alignment spectrum shows two clear clusters of topics, one retaining alignment levels comparable to the previous period, and the other exhibiting markedly low alignment. As discussed in Appendix~\ref{a:results-finland}, 
these patterns reflect the collaborative dynamics explicitly adopted to face the crisis.

\section{Case Study: Higher-Order Partisan Sorting of the U.S. Public}
    \label{s:casestudy}

\begin{figure}
    \centering
        \includegraphics[width=\textwidth]{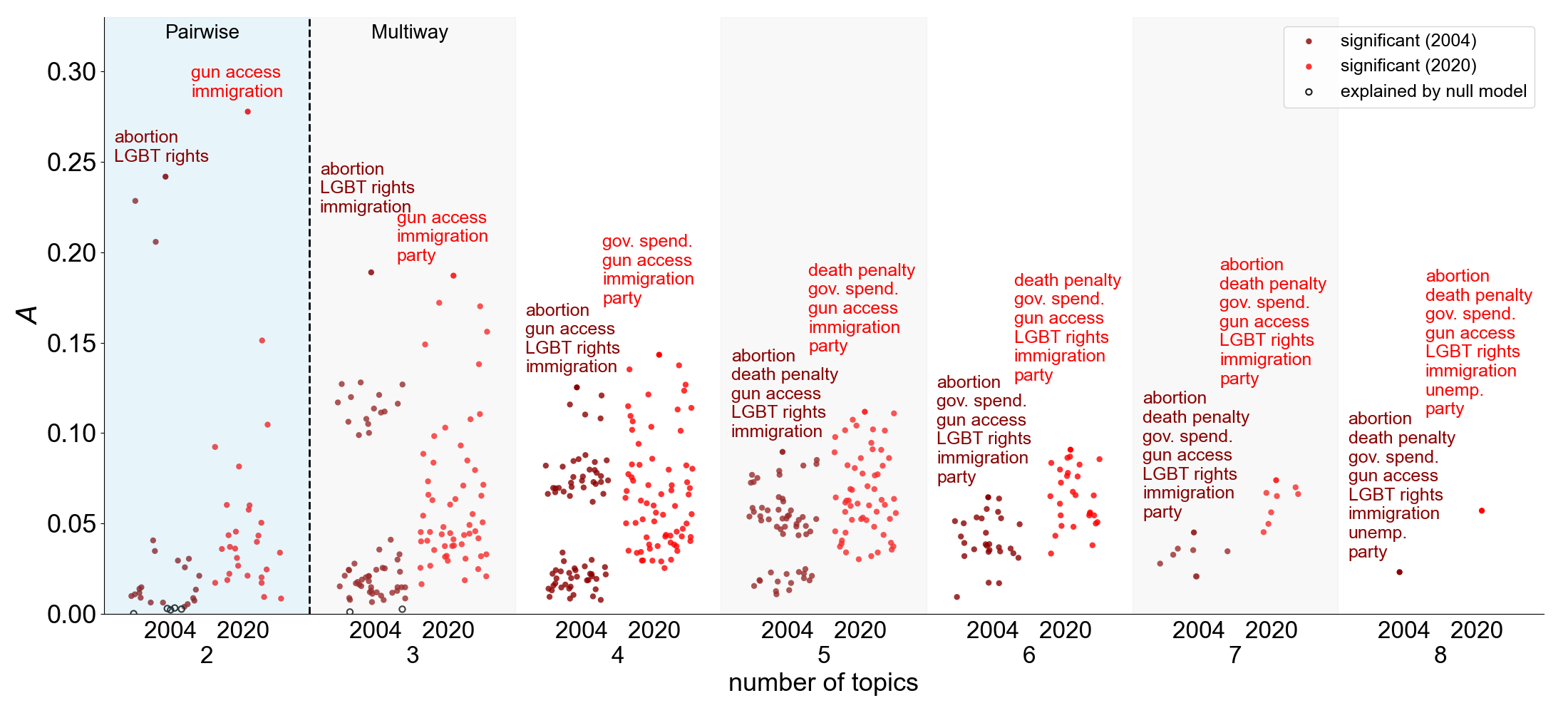}
    \caption{A comparison of multiway alignment between ANES Time Series data 2004 and 2020.
    }
    \label{fig:surveys_2004_2020}
\end{figure}

Building on concepts developed in the previous sections, we now turn our attention to a practical application through a case study that focuses on the alignment of political attitudes in the U.S. public. By analyzing the multiway alignment scores across multiple political topics, we aim to characterize partisan sorting in the United States. Our analysis uses data from the American National Election Studies (ANES) surveys to understand how political attitudes align across various topical issues. 

Established in 1948, the ANES surveys are conducted before and after every presidential election to investigate voter attitudes, party identification, and electoral behavior over time~\citep{anes}. ANES surveys have been used to study polarization, particularly affective polarization, by applying traditional statistical methods such as factor analysis~\citep{abramowitz2021s}, Cohen’s d coefficient~\citep{cavari2018polarized}, or regression models~\citep{enders2021value, lupton2020values, robison2019group} to identify underlying dimensions of political attitudes that contribute to polarization, measure the effect size of differences in political attitudes between groups, and quantify the relationships between demographic factors and the degree of polarization among the electorate.

Traditionally, studies on public opinion polarization have focused on individual issue attitudes, often overlooking how these attitudes are interconnected across multiple issues. This approach misses the broader picture of how people's opinions form a cohesive ideological framework and potentially sort individuals into opposing factions. As \citet{baldassarri2008partisans} argue, opinion alignment, rather than just opinion radicalization, affects social integration and political stability. Similarly, other authors have argued that political polarization is driven by social group identification and outgroup aversion~\citep{bougher2017correlates, mason2018uncivil}. Pursuing this idea, many authors have used ANES surveys to study how individuals' attitudes toward various issues align with party identification or with other issues~\citep{baldassarri2008partisans,kozlowski2021issuecorrelation, bougher2017correlates, hetherington2009polarization, luders2024attitude}. Still, the study of alignment has mostly been approached by measuring pairwise correlation between pairs of issues or between each issue and party identification~\citep{baldassarri2008partisans, kozlowski2021issuecorrelation, hetherington2009polarization}.

An attempt to consider multiple issues at the same time can be found in~\citet{bougher2017correlates}, where partisan alignment is measured by considering whether individuals' stances on an issue align with the opposing party's ideological position, with the own party's position, or with neither one. \citet{luders2024attitude} adopt a network-based approach to study attitude alignment over eight issues, constructing and visualizing a graph where nodes are stances and edges are weighted based on phi correlation coefficients between each pair of stances. While these studies focus on the organization of public opinion around ideologically- or partisan-opposed factions
that contain multiple attitudes, neither explicitly examine higher-order dependencies among the attitudes.

In this case study, we aim to characterize partisan-ideological sorting in terms of the topical issues which ideological groups coalesce their identities around. To do so, we analyze multiway alignment (both issue alignment and partisan alignment) in U.S. public opinion using the ANES Time Series data from 2004--2020. Instead of directly measuring polarization, we examine how the alignment of opinions across various issues gives rise to ideological groups and partisanship. 

Our analysis reveals several key findings. First, we observe an increase in multiway alignment over time, indicating that the interdependencies among the topics in analysis (seven issues and partisan identification) have intensified. This trend is shown in Figure~\ref{fig:surveys_2004_2020}, where we compare the full spectrum of the multiway alignment computed from the 2004 and 2020 data. While Figure~\ref{fig:surveys_2004_2020} suggests a more polarized landscape in contemporary American discourse, we also find that the alignment of stances concerning abortion and LGBT rights, which was predominant in 2004, has been replaced by a more pronounced alignment between gun access and immigration in 2020. Notably, the alignment score pertaining to the latter pair surpasses the three highest scores observed in 2004, signifying the existence of stronger 2-way constraints and the changing salience of different issues over time. Moreover, our results show that the overall multiway alignment across all eight topics has increased, portraying a higher level of attitude constraint in the public opinion.

The second key finding concerns partisan sorting. The analysis of 3-tuples and higher illustrates the growing importance of political partisanship in U.S. public opinion. Whereas in 2004, positions toward immigration played a central role within triadic-and-higher relationships, in 2020, party preferences became one of the most prominent topics, marking a shift towards sociopolitical affiliations as key alignment determinants. In this respect, our measure of multiway alignment allows us to investigate research questions that study partisanship across a multitude of issues. Specifically, we are now able to quantify the strength of the relationship between party preference and a bundle of multiple topics. 

\begin{figure}
    \centering
    \includegraphics[width=\linewidth,
    ]{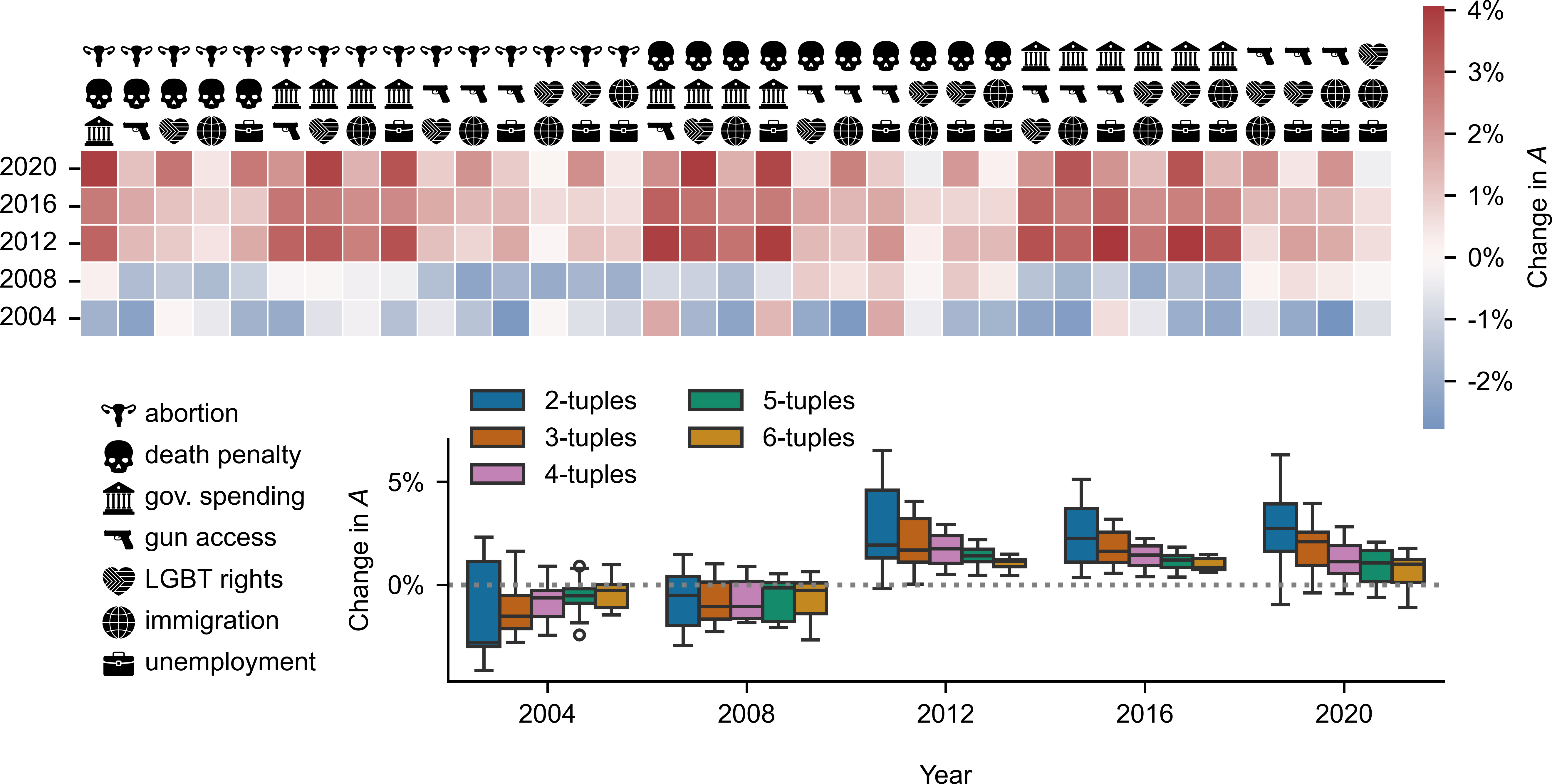}
\caption{The heatmap shows the percentage change of multiway alignment when adding party preference to any $3$-tuple of topics selected from ANES Time Series Study~\citep{anes}. The color gradient shows the increasing significance of party preference over time in $4$-way alignment. The timeseries chart in the bottom right shows how the same trend holds for tuples of any size.}
    \label{fig:anes-timeseries-triplets}
\end{figure}

We show this in the heatmap in Figure~\ref{fig:anes-timeseries-triplets}, which specifically illustrates the influence of including party preference into any 3-tuple formation as the percentage change in multiway alignment score. Notably, starting from 2012, the 4-way alignment of any four topics that include individual party preferences consistently surpasses the corresponding alignment scores obtained from the three topics other than party preference. This phenomenon is not restricted to 3-tuples: the relative change of the multiway alignment score when incorporating individual party preference within $k$-tuples ($k \in \{2, \dots, 6\}$) over a more complete range of years shows a visible trend (Figure~\ref{fig:anes-timeseries-triplets}), demonstrating an increasing alignment over time between party preferences and the remaining stances, across all values of $k$. The observed development of multiway alignment confirms the growing entwinement between political affiliations and multiple issues, suggesting that stances on various issues are increasingly catalyzed by political affiliations. These results not only demonstrate an increasing trend in partisan alignment across the general public of the U.S., but also identify the period between 2008 and 2012 as a pivotal point where the multi-dimensional constraint between party affiliation and the set of issues considered in this study becomes clearly significant.

\section{Discussion}
\label{s:discussion}

In this paper, we introduce a quantitative method for analyzing political alignment across multiple topics, building upon existing approaches that focus on pairwise alignment. Our multiway alignment measure retains similar interpretability as traditional pairwise approaches while offering a more comprehensive view of alignment patterns in various socio-political contexts. Our method proves especially useful for investigating both issue and partisan alignment across different systems, providing insights that go beyond what can be captured with pairwise methods alone.

Our approach can be applied in diverse contexts, including public opinion surveys, Parliamentary voting records, and online discussions. Our case study on ANES data demonstrates how our measure can be applied in contexts where opinion groups are straightforwardly defined by pre-existing categories, such as surveys responses. By applying our measure, we were able to detect and quantify the emergence of well-known partisan divides. In addition to survey data, our measure can also be applied to Parliamentary voting, where we can identify subsets of issues that exhibit high alignment based on co-voting patterns and irrespective of party affiliation. Another application of our measure is the analysis of social media discussions, where individual group membership is not explicitly available but can be inferred via clustering or community detection methods.

These use cases highlight the ability of our multiway alignment measure to uncover patterns of higher-order alignment that might not visible in pairwise analyses. We have highlighted this by presenting three archetypical shapes of the alignment spectrum, which visually demonstrate how positions on one topic can inform or constrain opinions on others. These patterns help clarify how ideological and partisan divisions manifest across multiple issues, supporting the study of alignment.

The implications of this work extend both theoretically and practically. Theoretically, the multiway alignment measure enhances our ability to study political polarization, not just at the level of individual issues, but with respect to the belief systems. This is particularly relevant in understanding which  issues or topics have the potential to drive polarization and societal divides. Furthermore, this approach enables research related to the complexity of ideological landscapes, allowing for more systematic assessments of how political identities and belief systems coalesce or fragment over time.

Practically, our approach equips researchers with a versatile tool for studying alignment across a range of contexts. However, some challenges remain. One limitation is the need for sufficiently large sample sizes, especially when analyzing many topics simultaneously. When the sample size is small relative to the number of topics, the results may be less robust. In such cases, we recommend comparing the alignment scores with those obtained from the null model described in Appendix~\ref{a:nullmodel}. Another challenge is the assumption that the data used to compute alignment is representative of broader societal attitudes, which may not always be the case, especially when relying on self-selecting platforms like online social media. Additionally, the computational complexity of calculating multiway alignment increases as the number of issues grows, which could pose practical constraints in large-scale datasets.

Despite these limitations, our method paves the way for several promising directions in future research. First, the multiway alignment measure could be applied in a comparative context, allowing researchers to explore whether the patterns observed in one political system hold across different systems or countries. Moreover, our measure can be used to study the evolution of alignment over time, particularly in response to major events like elections, social movements, or crises such as the COVID-19 pandemic.

In conclusion, our multiway alignment measure provides a powerful tool for understanding the structure of political attitudes across multiple dimensions. By uncovering higher-order alignment patterns, it offers new insights into the dynamics of ideological coherence, partisan divides, and issue-based polarization. While our case studies demonstrate the utility of the measure in specific contexts, its potential applications extend beyond the examples provided in this paper. We hope that this work sets the stage for future research on societal polarization and partisan animosity. To facilitate this, we offer an open-source Python package, \texttt{multiway\_alignment}, enabling researchers across disciplines to adopt and extend this methodology.

\paragraph{Acknowledgement}
We thank Boyoon Lee and Kevin Reuning for helpful feedback.

\paragraph{Funding Statement}

This research was supported by grants from the Research Council of Finland
(349366; 352561; 357743)

\paragraph{Competing Interests}

None.

\paragraph{Data Availability Statement}
Replication code for this article is available at \url{https://github.com/letiziaia/multiway-alignment}. All data used in this study is available publicly. For details, see Appendix~\ref{s:data} and Appendix~\ref{s:clustering}.

\bibliographystyle{chicago}
\bibliography{main}  

\begin{thebibliography}{}

\bibitem[\protect\citeauthoryear{??}{usd}{2021}]{usdata}
 (2021).
\newblock Voteview: Congressional roll-call votes database.

\bibitem[\protect\citeauthoryear{??}{edu}{2022}]{eduskunta_avoin_data_2022}
 (2022).
\newblock Eduskunta - avoin data.

\bibitem[\protect\citeauthoryear{Abdallah and Plumbley}{Abdallah and Plumbley}{2012}]{abdallah2012binding}
Abdallah, S.~A. and M.~D. Plumbley (2012).
\newblock A measure of statistical complexity based on predictive information with application to finite spin systems.
\newblock {\em Physics Letters A\/}~{\em 376\/}(4), 275--281.

\bibitem[\protect\citeauthoryear{Abramowitz}{Abramowitz}{2021}]{abramowitz2021s}
Abramowitz, A.~I. (2021).
\newblock It’s only you and me and we just disagree: The ideological foundations of affective polarization.
\newblock In {\em The Forum}, Volume~19, pp.\  349--364. De Gruyter.

\bibitem[\protect\citeauthoryear{Adler and Wilkerson}{Adler and Wilkerson}{2017}]{ustopics}
Adler, E.~S. and J.~Wilkerson (2017).
\newblock Congressional bills project.

\bibitem[\protect\citeauthoryear{{American National Election Studies}}{{American National Election Studies}}{2021}]{anes}
{American National Election Studies} (2021).
\newblock Anes 2020 time series study full release.

\bibitem[\protect\citeauthoryear{Bafumi and Shapiro}{Bafumi and Shapiro}{2009}]{bafumi2009new}
Bafumi, J. and R.~Y. Shapiro (2009).
\newblock A new partisan voter.
\newblock {\em The journal of politics\/}~{\em 71\/}(1), 1--24.

\bibitem[\protect\citeauthoryear{Baldassarri and Gelman}{Baldassarri and Gelman}{2008}]{baldassarri2008partisans}
Baldassarri, D. and A.~Gelman (2008).
\newblock Partisans without constraint: Political polarization and trends in american public opinion.
\newblock {\em American Journal of Sociology\/}~{\em 114\/}(2), 408--446.

\bibitem[\protect\citeauthoryear{Battiston, Cencetti, Iacopini, Latora, Lucas, Patania, Young, and Petri}{Battiston et~al.}{2020}]{battiston2020networks}
Battiston, F., G.~Cencetti, I.~Iacopini, V.~Latora, M.~Lucas, A.~Patania, J.-G. Young, and G.~Petri (2020).
\newblock Networks beyond pairwise interactions: Structure and dynamics.
\newblock {\em Physics Reports\/}~{\em 874}, 1--92.
\newblock Networks beyond pairwise interactions: Structure and dynamics.

\bibitem[\protect\citeauthoryear{Bell}{Bell}{2003}]{bell2003coinfo}
Bell, A.~J. (2003).
\newblock The co-information lattice.
\newblock In {\em Proceedings of the fifth international workshop on independent component analysis and blind signal separation: ICA}, Volume 2003.

\bibitem[\protect\citeauthoryear{Bougher}{Bougher}{2017}]{bougher2017correlates}
Bougher, L.~D. (2017).
\newblock The correlates of discord: identity, issue alignment, and political hostility in polarized america.
\newblock {\em Political Behavior\/}~{\em 39}, 731--762.

\bibitem[\protect\citeauthoryear{Brabec}{Brabec}{2020}]{brabec2020covotingnet}
Brabec, D. (2020).
\newblock The disintegration of kdu-{\v{c}}sl in 2009: The network analysis of co-voting strategies of the kdu-{\v{c}}sl deputies.
\newblock {\em Politics in Central Europe\/}~{\em 16\/}(2), 547--563.

\bibitem[\protect\citeauthoryear{Cavari and Freedman}{Cavari and Freedman}{2018}]{cavari2018polarized}
Cavari, A. and G.~Freedman (2018).
\newblock Polarized mass or polarized few? assessing the parallel rise of survey nonresponse and measures of polarization.
\newblock {\em The Journal of Politics\/}~{\em 80\/}(2), 719--725.

\bibitem[\protect\citeauthoryear{Chen, Salloum, Gronow, Yl{\"a}-Anttila, and Kivel{\"a}}{Chen et~al.}{2021}]{chen2021polarization}
Chen, T. H.~Y., A.~Salloum, A.~Gronow, T.~Yl{\"a}-Anttila, and M.~Kivel{\"a} (2021).
\newblock Polarization of climate politics results from partisan sorting: Evidence from finnish twittersphere.
\newblock {\em Global Environmental Change\/}~{\em 71}, 102348.

\bibitem[\protect\citeauthoryear{Conover, Ratkiewicz, Francisco, Gon{\c{c}}alves, Menczer, and Flammini}{Conover et~al.}{2011}]{conover2011polarizationtwitter}
Conover, M., J.~Ratkiewicz, M.~Francisco, B.~Gon{\c{c}}alves, F.~Menczer, and A.~Flammini (2011).
\newblock Political polarization on twitter.
\newblock In {\em Proceedings of the international aaai conference on web and social media}, Volume~5, pp.\  89--96.

\bibitem[\protect\citeauthoryear{Converse}{Converse}{2006}]{converse2006nature}
Converse, P.~E. (2006).
\newblock The nature of belief systems in mass publics (1964).
\newblock {\em Critical review\/}~{\em 18\/}(1-3), 1--74.

\bibitem[\protect\citeauthoryear{Davies and Bouldin}{Davies and Bouldin}{1979}]{daviesbouldin}
Davies, D.~L. and D.~W. Bouldin (1979).
\newblock A cluster separation measure.
\newblock {\em IEEE transactions on pattern analysis and machine intelligence\/}~(2), 224--227.

\bibitem[\protect\citeauthoryear{Eduskunta.fi}{Eduskunta.fi}{2020}]{eduskuntapressrelease}
Eduskunta.fi (2020, 3).
\newblock Perussuomalaiset vetää pois välikysymyksen koronatilanteen takia.

\bibitem[\protect\citeauthoryear{Enders and Lupton}{Enders and Lupton}{2021}]{enders2021value}
Enders, A.~M. and R.~N. Lupton (2021).
\newblock Value extremity contributes to affective polarization in the us.
\newblock {\em Political Science Research and Methods\/}~{\em 9\/}(4), 857--866.

\bibitem[\protect\citeauthoryear{Ester, Kriegel, Sander, Xu, et~al.}{Ester et~al.}{1996}]{dbscan}
Ester, M., H.-P. Kriegel, J.~Sander, X.~Xu, et~al. (1996).
\newblock A density-based algorithm for discovering clusters in large spatial databases with noise.
\newblock In {\em kdd}, Volume~96, pp.\  226--231.

\bibitem[\protect\citeauthoryear{Fishman and Davis}{Fishman and Davis}{2022}]{fishman2022change}
Fishman, N. and N.~T. Davis (2022).
\newblock Change we can believe in: Structural and content dynamics within belief networks.
\newblock {\em American Journal of Political Science\/}~{\em 66\/}(3), 648--663.

\bibitem[\protect\citeauthoryear{Flores, Cole, Dickert, Eom, Jiga-Boy, Kogut, Loria, Mayorga, Pedersen, Pereira, et~al.}{Flores et~al.}{2022}]{flores2022politicians}
Flores, A., J.~C. Cole, S.~Dickert, K.~Eom, G.~M. Jiga-Boy, T.~Kogut, R.~Loria, M.~Mayorga, E.~J. Pedersen, B.~Pereira, et~al. (2022).
\newblock Politicians polarize and experts depolarize public support for covid-19 management policies across countries.
\newblock {\em Proceedings of the National Academy of Sciences\/}~{\em 119\/}(3), e2117543119.

\bibitem[\protect\citeauthoryear{Garimella, Morales, Gionis, and Mathioudakis}{Garimella et~al.}{2018}]{garimella2018quantifying}
Garimella, K., G.~D.~F. Morales, A.~Gionis, and M.~Mathioudakis (2018).
\newblock Quantifying controversy on social media.
\newblock {\em ACM Transactions on Social Computing\/}~{\em 1\/}(1), 1--27.

\bibitem[\protect\citeauthoryear{Gibbons and Kendall}{Gibbons and Kendall}{1990}]{gibbons1990rank}
Gibbons, J.~D. and M.~Kendall (1990).
\newblock Rank correlation methods.
\newblock {\em Edward Arnold\/}~{\em 46}.

\bibitem[\protect\citeauthoryear{Hetherington}{Hetherington}{2009}]{hetherington2009polarization}
Hetherington, M.~J. (2009).
\newblock Putting polarization in perspective.
\newblock {\em British Journal of Political Science\/}~{\em 39\/}(2), 413--448.

\bibitem[\protect\citeauthoryear{Iannucci, Faqeeh, Salloum, Chen, and Kivelä}{Iannucci et~al.}{2024}]{ourtwitterdata}
Iannucci, L., A.~Faqeeh, A.~Salloum, T.~H.~Y. Chen, and M.~Kivelä (2024).
\newblock Multiway alignment of twitter networks from 2019 and 2023 finnish parliamentary elections [data set].

\bibitem[\protect\citeauthoryear{James, Ellison, and Crutchfield}{James et~al.}{2011}]{james2011enigm}
James, R.~G., C.~J. Ellison, and J.~P. Crutchfield (2011).
\newblock Anatomy of a bit: Information in a time series observation.
\newblock {\em Chaos: An Interdisciplinary Journal of Nonlinear Science\/}~{\em 21\/}(3).

\bibitem[\protect\citeauthoryear{Karypis and Kumar}{Karypis and Kumar}{1998}]{metis}
Karypis, G. and V.~Kumar (1998).
\newblock A fast and high quality multilevel scheme for partitioning irregular graphs.
\newblock {\em SIAM Journal on scientific Computing\/}~{\em 20\/}(1), 359--392.

\bibitem[\protect\citeauthoryear{Kozlowski and Murphy}{Kozlowski and Murphy}{2021}]{kozlowski2021issuecorrelation}
Kozlowski, A.~C. and J.~P. Murphy (2021).
\newblock Issue alignment and partisanship in the american public: Revisiting the ‘partisans without constraint’ thesis.
\newblock {\em Social Science Research\/}~{\em 94}, 102498.

\bibitem[\protect\citeauthoryear{Lehtonen and Yl{\"a}-Anttila}{Lehtonen and Yl{\"a}-Anttila}{2024}]{lehtonen2024navigating}
Lehtonen, J. and T.~Yl{\"a}-Anttila (2024).
\newblock Navigating pandemic waves: Consensus, polarisation and pluralism in the finnish parliament during covid-19.
\newblock {\em Politics\/}, 02633957241259085.

\bibitem[\protect\citeauthoryear{Levin, Milner, and Perrings}{Levin et~al.}{2021}]{levin2021dynamics}
Levin, S.~A., H.~V. Milner, and C.~Perrings (2021).
\newblock The dynamics of political polarization.
\newblock {\em Proceedings of the National Academy of Sciences\/}~{\em 118\/}(50), e2116950118.

\bibitem[\protect\citeauthoryear{L{\"u}ders, Carpentras, and Quayle}{L{\"u}ders et~al.}{2024}]{luders2024attitude}
L{\"u}ders, A., D.~Carpentras, and M.~Quayle (2024).
\newblock Attitude networks as intergroup realities: Using network-modelling to research attitude-identity relationships in polarized political contexts.
\newblock {\em British Journal of Social Psychology\/}~{\em 63\/}(1), 37--51.

\bibitem[\protect\citeauthoryear{Lupton, Smallpage, and Enders}{Lupton et~al.}{2020}]{lupton2020values}
Lupton, R.~N., S.~M. Smallpage, and A.~M. Enders (2020).
\newblock Values and political predispositions in the age of polarization: Examining the relationship between partisanship and ideology in the united states, 1988--2012.
\newblock {\em British Journal of Political Science\/}~{\em 50\/}(1), 241--260.

\bibitem[\protect\citeauthoryear{Mason}{Mason}{2018}]{mason2018uncivil}
Mason, L. (2018).
\newblock {\em Uncivil agreement: How politics became our identity}.
\newblock University of Chicago Press.

\bibitem[\protect\citeauthoryear{Meil{\u{a}}}{Meil{\u{a}}}{2003}]{variation-info}
Meil{\u{a}}, M. (2003).
\newblock Comparing clusterings by the variation of information.
\newblock In {\em Learning Theory and Kernel Machines: 16th Annual Conference on Learning Theory and 7th Kernel Workshop, COLT/Kernel 2003, Washington, DC, USA, August 24-27, 2003. Proceedings}, pp.\  173--187. Springer.

\bibitem[\protect\citeauthoryear{Newman, Cantwell, and Young}{Newman et~al.}{2020}]{reduced-mi}
Newman, M.~E., G.~T. Cantwell, and J.-G. Young (2020).
\newblock Improved mutual information measure for clustering, classification, and community detection.
\newblock {\em Physical Review E\/}~{\em 101\/}(4), 042304.

\bibitem[\protect\citeauthoryear{Niemikari and Raunio}{Niemikari and Raunio}{2022}]{niemikari2022centralized}
Niemikari, R. and T.~Raunio (2022).
\newblock Centralized leadership, ministerial dominance, and improvised instruments: The governance of covid in finland.
\newblock {\em Nordisk Administrativt Tidsskrift\/}~{\em 99\/}(2), 1–--19.

\bibitem[\protect\citeauthoryear{Robison and Moskowitz}{Robison and Moskowitz}{2019}]{robison2019group}
Robison, J. and R.~L. Moskowitz (2019).
\newblock The group basis of partisan affective polarization.
\newblock {\em The Journal of Politics\/}~{\em 81\/}(3), 1075--1079.

\bibitem[\protect\citeauthoryear{Rosas, Mediano, Gastpar, and Jensen}{Rosas et~al.}{2019}]{rosas2019oinfo}
Rosas, F.~E., P.~A. Mediano, M.~Gastpar, and H.~J. Jensen (2019).
\newblock Quantifying high-order interdependencies via multivariate extensions of the mutual information.
\newblock {\em Physical Review E\/}~{\em 100\/}(3), 032305.

\bibitem[\protect\citeauthoryear{Rosenberg and Hirschberg}{Rosenberg and Hirschberg}{2007}]{homogeneity}
Rosenberg, A. and J.~Hirschberg (2007).
\newblock V-measure: A conditional entropy-based external cluster evaluation measure.
\newblock In {\em Proceedings of the 2007 joint conference on empirical methods in natural language processing and computational natural language learning (EMNLP-CoNLL)}, pp.\  410--420.

\bibitem[\protect\citeauthoryear{Rousseeuw}{Rousseeuw}{1987}]{silhouettes}
Rousseeuw, P.~J. (1987).
\newblock Silhouettes: a graphical aid to the interpretation and validation of cluster analysis.
\newblock {\em Journal of computational and applied mathematics\/}~{\em 20}, 53--65.

\bibitem[\protect\citeauthoryear{Salloum, Chen, and Kivel{\"a}}{Salloum et~al.}{2022}]{alinormalization}
Salloum, A., T.~H.~Y. Chen, and M.~Kivel{\"a} (2022).
\newblock Separating polarization from noise: comparison and normalization of structural polarization measures.
\newblock {\em Proceedings of the ACM on human-computer interaction\/}~{\em 6\/}(CSCW1), 1--33.

\bibitem[\protect\citeauthoryear{Salloum, Chen, and Kivelä}{Salloum et~al.}{2024}]{salloum2024anatomy}
Salloum, A., T.~H.~Y. Chen, and M.~Kivelä (2024).
\newblock Anatomy of elite and mass polarization in social networks.
\newblock arXiv:2406.12525.

\bibitem[\protect\citeauthoryear{Sun}{Sun}{1975}]{sun1975dtc}
Sun, T. (1975).
\newblock Linear dependence structure of the entropy space.
\newblock {\em Inf Control\/}~{\em 29\/}(4), 337--68.

\bibitem[\protect\citeauthoryear{Van~Bavel and Pereira}{Van~Bavel and Pereira}{2018}]{van2018partisan}
Van~Bavel, J.~J. and A.~Pereira (2018).
\newblock The partisan brain: An identity-based model of political belief.
\newblock {\em Trends in cognitive sciences\/}~{\em 22\/}(3), 213--224.

\bibitem[\protect\citeauthoryear{Vinh, Epps, and Bailey}{Vinh et~al.}{2009}]{nmi}
Vinh, N.~X., J.~Epps, and J.~Bailey (2009).
\newblock Information theoretic measures for clusterings comparison: is a correction for chance necessary?
\newblock In {\em Proceedings of the 26th annual international conference on machine learning}, pp.\  1073--1080.

\bibitem[\protect\citeauthoryear{Watanabe}{Watanabe}{1960}]{watanabe1960tc}
Watanabe, S. (1960).
\newblock Information theoretical analysis of multivariate correlation.
\newblock {\em IBM Journal of research and development\/}~{\em 4\/}(1), 66--82.

\bibitem[\protect\citeauthoryear{Westwood, Iyengar, Walgrave, Leonisio, Miller, and Strijbis}{Westwood et~al.}{2018}]{westwood2018tie}
Westwood, S.~J., S.~Iyengar, S.~Walgrave, R.~Leonisio, L.~Miller, and O.~Strijbis (2018).
\newblock The tie that divides: Cross-national evidence of the primacy of partyism.
\newblock {\em European Journal of Political Research\/}~{\em 57\/}(2), 333--354.

\bibitem[\protect\citeauthoryear{YLE}{YLE}{2020a}]{YLE2020a}
YLE (2020a, 3).
\newblock Finland closes schools, declares state of emergency over coronavirus.

\bibitem[\protect\citeauthoryear{YLE}{YLE}{2020b}]{YLE2020b}
YLE (2020b, 3).
\newblock Finland shuts down uusimaa to fight coronavirus.

\end{thebibliography}

\appendix
\section*{Appendix}

\section{Data}
    \label{s:data}

We operationalize the suggested multiway alignment framework across diverse datasets, including survey data%(described in~\ref{ss:surveydata})
, retweet networks% (\ref{ss:twitterdata})
, and Parliamentary roll-call votes.% (\ref{ss:rollcalldata}).

\subsection{Survey Data}
    \label{ss:surveydata}

The survey data are sourced from the ANES 2020 Time Series Study, encompassing interview responses from eligible U.S. voters across various years. The interview questions span a spectrum of issues, including topics such as public health, faith in experts and science, group empathy, and societal concerns. From the available range of topics, we selected questions about immigration (\texttt{VCF9223}), abortion (\texttt{VCF0838}), gun access (\texttt{VCF9238}), unemployment (\texttt{VCF9229}), party preference (\texttt{VCF9205}), death penalty (\texttt{VCF9236}), LGBT rights (\texttt{VCF0876}), government spending (\texttt{VCF0839}). This selection facilitates an examination of the evolution of multi-issue alignment across the years 2004 to 2020.

\subsection{Retweet Networks}
    \label{ss:twitterdata}

In line with the methodology employed in \citet{chen2021polarization}, we apply our algorithm to Twitter data related to Finnish political elections in 2019. Additionally, we extend our analysis to include data from the Finnish political elections in 2023 \citep{salloum2024anatomy}.

For the 2019 dataset, tweets and retweets were gathered during the same period (March to July 2019), using a predefined set of Finnish hashtags as in~\citet{chen2021polarization}. The collected data were then categorized into 11 distinct labels based on their text content. These labels encompass conversation streams related to Finnish political parties (SDP, National Coalition, Finns Party, Green Party, Left Party, Center Party) and salient topics (economic policy, social security, immigration, climate, education). Given that a single tweet can relate to multiple categories simultaneously, this approach results in a multilabel classification. Subsequently, for each label, we constructed the corresponding retweet network as detailed in~\ref{ss:retweet-clust}. 

The dataset related to the 2023 Finnish elections comprises tweets and retweets published from September to April 2023, and was collected by using a similar set of Finnish hashtags as for 2019~\citep{chen2021polarization}. Similar to the 2019 data, we categorized tweets and retweets into the same 11 labels and subsequently build the retweet networks by following the same approach as for 2019.

\subsection{Roll-Call Data}
    \label{ss:rollcalldata}

The multiway approach described in the paper has been applied to voting data in Finnish Parliament and U.S. House. Data sources and data description are provided below.

\subsubsection{Finnish Parliament}
    \label{sss:finland}
\begin{figure*}
    \centering
    \includegraphics[width=1\textwidth]{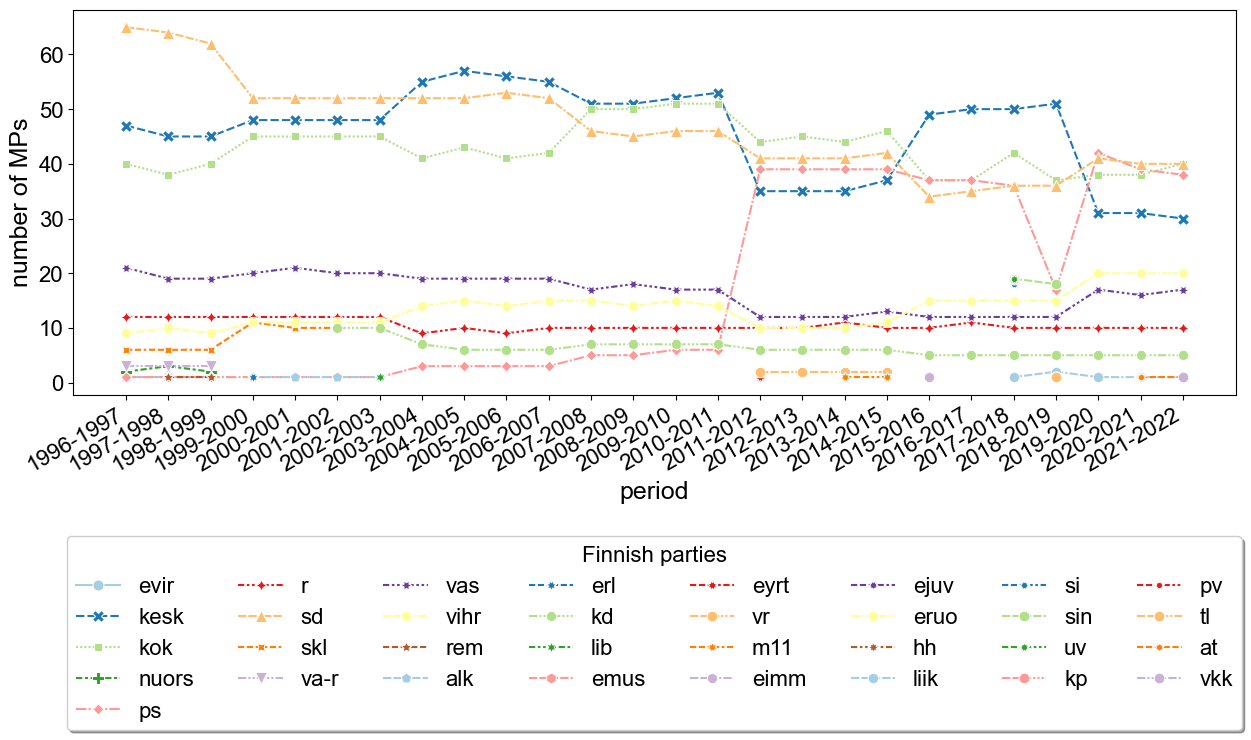}
    \caption{Number of Parliament Members (MPs) by party in the Finnish Parliament.}
    \label{fig:fin_parties}
\end{figure*}

The dataset of Parliamentary activities has been collected via the REST API offered by the Finnish government~\citep{eduskunta_avoin_data_2022}. The data is available both in Finnish and in Swedish, but policy topics are not explicitly given. As the analysis has been conducted on the Finnish version of the data, the Finnish titles of the bills have been used to derive the policy labels. In order to do so, Finnish words have been stemmed and stop-words have been removed. Then, we obtain a feature matrix with latent semantic analysis (LSA), by applying truncated singular value decomposition (SVD) to the term frequency-inverse document frequency (tf-idf) matrix. Finally, major topics are found by K-Means clustering, where the number of clusters is chosen by minimizing Davies-Bouldin score~\citep{daviesbouldin}. The Finnish version of the data from~\citet{eduskunta_avoin_data_2022} indicates each Parliament member's vote by utilizing one among \textit{jaa} (yes), \textit{ei} (no), \textit{poissa} (absent), \textit{tyhjä} (empty). For the further analysis, the voting behaviour is simplified to \textit{yes}, \textit{no}, and \textit{other}, where \textit{yes} and \textit{no} correspond exactly to the original categories, whereas the category \textit{other} covers both \textit{poissa} and \textit{tyhjä}. Among the 33 different parties that appear in the Finnish Parliament between 1996 and 2021 (Figure~\ref{fig:fin_parties}), the 8 biggest parties were used for the analysis as they appear in all years and account for the vast majority of the Parliament. Specifically, the 8 parties included in the analysis are \textit{Centre Party} (\texttt{kesk}), \textit{National Coalition Party} (\texttt{kok}),  \textit{Finns Party} (\texttt{ps}),  \textit{Swedish People's Party of Finland} (\texttt{r}), \textit{Social Democratic Party} (\texttt{sd}), \textit{Left Alliance} (\texttt{vas}), \textit{Green League} (\texttt{vihr}), \textit{Christian Democrats} (\texttt{kd}). For the following analysis, each Parliamentary period is divided into 4 approximately equal-sized sub-periods.

\subsubsection{U.S. House}
    \label{sss:usa}

A complete dataset covering all congresses from 1 to 117 and both chambers (\textit{House} and \textit{Senate}) is available from Voteview~\citep{usdata}. As for policy topics, the ones provided by Comparative Agendas~\citep{ustopics} for congresses between 97 and 113 are linked to the corresponding votes. Since the Independent party (party code \texttt{328}) has very limited number of members over the Congresses, it has been omitted from the analysis. As a result, the findings describe the dynamics of members of Republican Party and Democratic Party only. Furthermore, each Congress is split into 2 approximately equal-sized sub-periods, roughly corresponding to calendar years.

\newpage
\section{Opinion Partitions from Clustering or Community Detection}
\label{s:clustering}

Opinion groups have been extensively studied using clustering and community detection techniques~\citep{conover2011polarizationtwitter, chen2021polarization, brabec2020covotingnet}. These are powerful techniques for uncovering patterns in data, identifying distinct subgroups based on similarities in their actions or beliefs, and shedding light on the dynamics of polarization or on the diffusion of ideas within a society or network. Taking a similar approach, the idea of clusters or communities can be harnessed to investigate opinion groups across multiple issues simultaneously.

\subsection{Parliamentary Roll-Calls}
    \label{ss:rollcalls}

In a roll-call vote, each member of the Parliament can vote either yes or no or can abstain. However, for each policy area, there are typically multiple roll-call votes. Therefore, for each policy area, we can infer the opinion groups from the similarity in the vector of votes each individual MP has casted on all the roll-calls regarding a specific policy domain.

For each issue or policy domain, each of the Parliament members (MPs) is represented with the vector of votes they have casted on all the roll-calls regarding the specific issue. Specifically, for each issue, each MP is associated to a vector $u_{i}=(u_{i1}, u_{i2}, ...)$, where the element $u_{ij}$ is the vote casted by the MP $i$ on the $j$-th roll-call regarding that issue, and can take values $-1$ (\textit{no}), $1$ (\textit{yes}), or $0$ (\textit{other}, as described in~\ref{ss:rollcalldata}). This representation allows us to detect MPs with highly co-voting behavior from the cosine similarity $u_{m}.u_{n}$ between their vectors. In fact, Parliament members who cast the same votes have small cosine distance and high cosine similarity. As a result, for each issue, we can build the similarity network of the members of Parliament, where the interaction between each pair of MPs is modeled by a link weighted proportional to the similarity of their voting behavior. More precisely, given one policy domain or issue, the network of Parliament members is defined as follows:
\begin{itemize}
    \item each MP is mapped to a node in the network;
    \item each pair of MPs is connected by a link that is weighted by the cosine similarity between the vectors of votes casted by the two MPs, where each vote is either $1$ (yes), $-1$ (no), or $0$ (has not voted).
\end{itemize}
By definition, the resulting network is an undirected weighted all-to-all network that carries information about co-voting behavior of the MPs on one specific issue. The remaining layers of the network are defined similarly, each time by considering the votes associated with the topical issue of interest.

Finally, we repeatedly apply clustering to each issue network independently, to find communities of strongly collaborating MPs at each layer. Specifically, the opinion partitions for each issue for Parliamentary roll-calls data are discovered via layer-level clustering performed with Density-Based Spatial Clustering of Applications with Noise (DBSCAN)~\citep{dbscan}, that detects areas of high density, separated by areas of lower density. Therefore, the algorithm neither requires to know the number of clusters beforehand nor assumes convexity of the clusters. The clusters are expanded from a number of core samples by recursively including their neighbours. On the other hand, samples that are distant from any core points are considered to be outliers. In practice, this means that, for each policy domain or topic $T_i$, the algorithm is able to find the optimal number of clusters and a number of outliers, depending on the co-voting behaviour on the particular issue.

The hyperparameters of DBSCAN are optimized based on the Silhouette score~\citep{silhouettes}, which compares the mean intra-cluster distance and the mean nearest-cluster distance to assess how well the clusters are separated. By applying DBSCAN to the pre-computed cosine distance matrix of voting behavior, we identify groups of highly collaborating MPs over each issue. Therefore, the labelings resulting from DBSCAN define the opinion partitions on which we apply the proposed multiway alignment measure.

\subsection{Retweet Networks}
    \label{ss:retweet-clust}

In each of the retweet networks, nodes represent users who shared a tweet (original or retweet) containing at least one relevant hashtag. Nodes are connected by an undirected edge if one user has retweeted the other on a given topic. Following the approach established in \citet{chen2021polarization} and \citet{garimella2018quantifying}, we employed the METIS algorithm~\citep{metis} to separately partition each networks into two similarly-sized groups with minimal connections between them. To maintain a realistic representation of group distribution in the network, a maximum imbalance constraint of 3:7 was applied, preventing the algorithm from producing a trivial partition. Partitions for each of the retweet networks are available in Zenodo~\citep{ourtwitterdata}.

\newpage
\section{Maximal Alignment Curve}
\label{a:maximal-alignment-curve}

The score defined in Equation~\ref{eq:general} can be computed for each $k$-combination of issues. For every value of $k$, our primary interest lies in the $k$-combination that maximizes $A$. To this scope, the maximal alignment curve summarizes the behavior of the system by considering the sets of issues exhibiting the highest alignment. In fact, the maximal alignment curve is a collection of the highest multiway alignment scores at each $k$.

As the maximal alignment curve is upper-bounded by $1$, the area under it describes the alignment of the entire system across all the issues. In fact, if the multiway alignment score remains close to $1$ across all $k$, it signifies that all issues exhibit extraordinarily high alignment. Conversely, the maximal alignment curve rapidly decreases if the consensus groups deteriorate, degenerating towards singletons as more issues are considered.

\begin{algorithm}
\footnotesize
    \SetKwInOut{Input}{Input}
    \SetKwInOut{Output}{Output}

    \underline{function getAlignmentCurve} $(M)$\;
    \Input{A 2-dimensional matrix $M$ of size $n \times m$, where $n$ is the number of individuals, $m$ is the number of issues, and $M_{i,j}$ is the label of the opinion group to which person $i$ belongs with respect to issue $j$}

    \Output{ getAlignmentCurve$(M)$, mapping each maximally aligned subset of issues of size $2, \dots m$ to its $A_{S_k}$ score}

    result = Map();

    \For{$k\leftarrow 2$ \KwTo $m$}{

        allCombinationsSizeK = getCombinations(issues, k);

        bestScore = 0;

        \For{$c\leftarrow 1$ \KwTo $m \choose k$}{

            issueSubset = allCombinationsSizeK[c];

            topicsOpinions = M[$1:$n][issueSubset];

            consensusPartition = getConsensusPartion(topicsOpinions);

            score = $0$;

            \For{$j\leftarrow 1$ \KwTo k}{

                score $+=$ S(consensusPartition, topicsOpinions[$1:$n][j]);

            }

            score $/=$ k;

            \If{score $>$ bestScore}{

                bestScore = score;

                bestKCombination = policies;

                }
        }

        result[bestKCombination] = bestScore;

     }

     return result;
    \caption{Maximal alignment curve}
    \label{algo:max_nmi}
\end{algorithm}

\newpage
\section{Interpretation of $A_{\text{MI}}$ Measure}
\label{a:mi_k3}

Given the set of $n$ topics $\mathcal{N} = \{ T_1, \ldots, T_n \}$, we show that, when the similarity measure $S$ is the mutual information (MI), the definition of multiway alignment given in Equation~\ref{eq:general} is equivalent to 

\begin{equation}\label{eq:mi}
A^{(n)} = \left ( \sum_{i=1}^n H(T_i) \right ) - H(T_1, \ldots, T_n) - \frac{1}{n} \left ( \sum_{s \in {\mathcal{N} \choose n - 1}} \sum_{k=2}^{n-1} \Bar{A_s}^{(k)} \right )
\end{equation}
where the terms $\Bar{A_s}$ are averages of lower-order multiway alignment over subsets of $\mathcal{N}$.

By our construction in Section~\ref{ss:multiway_alignment_score}, 

$$
A^{(n)} = \frac{1}{n} \sum_{i=1}^{n} \text{MI}(C(T_1, \ldots, T_{i-1}, T_{i+1}, \ldots, T_n), T_i),
$$
where MI is the pairwise mutual information, $C(T_1, \ldots, T_{i-1}, T_{i+1}, \ldots, T_n)$ is the consensus partition of individuals defined by the topics $T_1, \ldots, T_n$ excluding $T_i$, and each $T_i$ defines the opinion groups for the $i$-th individual topic. Therefore, for $n=2$, $A^{(2)} = MI(T_1, T_2) = \sum_{i=1}^2 H(T_i) - H(T_1, T_2)$. Similarly, for $n=3$, 

$$
A^{(3)} = \left ( \sum_{i=1}^3 H(T_i) \right ) - H(T_1, T_2, T_3) -  \frac{1}{3} \left ( {A^{(2)}}_{23} + {A^{(2)}}_{13} + {A^{(2)}}_{12} \right ),
$$
as $\Bar{A_{s}}^{(n-1)} = {A_{s}}^{(n-1)}$ for each $s \in {\mathcal{N} \choose n-1}$. 

Let us now assume that the statement in~\ref{eq:mi} is true for some arbitrary $n-1 \geq 2$, that is, 

$$A^{(n-1)} = \left ( \sum_{i=1}^{n-1} H(T_i) \right ) - H(T_1, \ldots, T_{n-1}) -  \frac{1}{n-1} \left ( \sum_{s \in {\mathcal{N} - 1 \choose n - 2}} \sum_{k=2}^{n-2} \Bar{A_s}^{(k)} \right )$$
and therefore 

$$
H(T_1, \ldots, T_{n-1}) =  \left ( \sum_{i=1}^{n-1} H(T_i) \right ) -  \frac{1}{n-1} \left ( \sum_{s \in {\mathcal{N} - 1 \choose n - 2}} \sum_{k=2}^{n-2} \Bar{A_s}^{(k)} \right ) -A^{(n-1)}.
$$

Then,

$$
A^{(n)} = \frac{1}{n} \sum_{i=1}^{n} \text{MI}(C(T_1, \ldots, T_{i-1}, T_{i+1}, \ldots, T_n), T_i)
$$

$$
= \frac{1}{n} \sum_{i=1}^{n} H(T_i) +  \frac{1}{n} \sum_{i=1}^{n} H(T_1, \ldots, T_{i-1}, T_{i+1}, \ldots, T_n) - H(T_1, \ldots, T_n)
$$

$$
= \frac{1}{n} \sum_{i=1}^{n} H(T_i) 
+  \frac{1}{n} 
    \sum_{i=1}^{n}
        \left ( 
            \sum_{j=1, j \neq i}^{n} H(T_j) 
        - \frac{1}{n-1} \left ( 
                \sum_{t \in {\mathcal{N}-1 \choose n-2}}
                    \sum_{k=2}^{n-2} \Bar{A_s}^{(k)} 
        \right )        
        - A^{(n-1)}
    \right )
- H(T_1, \ldots, T_n)
$$

$$
= \frac{1}{n} \sum_{i=1}^{n} H(T_i) 
+  \frac{1}{n} 
    \sum_{s\in {\mathcal{N} \choose n - 1}}
        \left ( 
            \sum_{j \in s} H(T_j) 
        - \frac{1}{n-1} \left ( 
                \sum_{t \in {s \choose n-2}}
                    \sum_{k=2}^{n-2} \Bar{A_t}^{(k)} 
        \right )        
        - {A_s}^{(n-1)}
    \right )
- H(T_1, \dots, T_n)
$$

$$
= \frac{1}{n} 
    \left (
        \sum_{i=1}^{n} H(T_i) 
        + \sum_{i=1}^{n}
            \sum_{j=1, j \neq i}^{n} H(T_j) 
    \right )
-  \frac{1}{n} 
    \sum_{s\in {\mathcal{N} \choose n - 1}}
        \left ( 
            \sum_{k=2}^{n-2} \Bar{A_s}^{(k)} 
        + {A_s}^{(n-1)}
    \right )
- H(T_1, \dots ,  T_n)
$$

$$
= \sum_{i=1}^{n} H(T_i) 
- \frac{1}{n} \left ( \sum_{s \in {\mathcal{N} \choose n - 1}} \sum_{k=2}^{n-1} \Bar{A_s}^{(k)} \right )
- H(T_1, \ldots, T_n),
$$
which also shows that each ${A_s}^{(k)}$ term contributes to the score by a fraction $\frac{1}{n \cdot (n-1) \cdot \ldots \cdot (k + 1)}$ as long as $3 \leq k + 1 \leq n$. For example, when $n=4$,

$$
A^{(4)} = \left ( \sum_{i=1}^4 H(T_i) \right ) - H(T_1, T_2, T_3, T_4) - \frac{1}{4}  \left ( A^{(3)}_{123} + A^{(3)}_{134} + A^{(3)}_{124} + A^{(3)}_{234} \right ) - \frac{1}{4} \left ( \Bar{A}^{(2)}_{123} + \Bar{A}^{(2)}_{124} + \Bar{A}^{(2)}_{134} + \Bar{A}^{(2)}_{234} \right )
$$
where

$$\Bar{A}^{(2)}_{123} =\frac{1}{3} \left ( A_{23} + A_{13} + A_{12} \right )$$
and each term of order $3$ contributes to the score by $\frac{1}{4}$. Therefore, the term

$$
\frac{1}{n} \left ( \sum_{s \in {\mathcal{N} \choose n - 1}} \sum_{k=2}^{n-1} \Bar{A_s}^{(k)} \right )
$$
in Equation~\ref{eq:mi} can also be understood as

\begin{equation}
\frac{1}{n}
        \sum_{s_1 \in {\mathcal{N} \choose n - 1}} 
            \left ( 
                {A_{s_1}}^{(n-1)} + \frac{1}{n-1}
                    \left ( 
                        \sum_{s_2 \in {s_1 \choose n - 2}} 
                            \left ( 
                                {A_{s_2}}^{(n-2)} + \frac{1}{n-2}
                                    \left (  
                                        \sum_{s_3 \in {s_2 \choose n - 3}} 
                                            \left ( 
                                                {A_{s_3}}^{(n-3)} + \frac{1}{n-3} 
                                                    \left (
                                                        \ldots 
                                                    \right )
                                            \right)
                                    \right )
                            \right )
                    \right )
            \right ).
\end{equation}

This average (\text{avg}) can be expressed recursively as:
\[
\text{avg}_{n} = \frac{1}{n} \sum_{s_1 \in {\mathcal{N} \choose n - 1}} 
            \left ( 
                {A_{s_1}}^{(n-1)} + \text{avg}_{n-1}(s_1)
            \right ),
\]
where
\[
\text{avg}_{n-k}(s_{k}) = \frac{1}{n-k} {A^{(n-k-1)}_{s_{k}}}
\]
when $n-k = 3$ and $n-k-1 = 2$.

\newpage
\section{Relation to Other Multivariate Measures}
\label{a:other-measures}

Co-information~\citep{bell2003coinfo} quantifies the amount of information gained about one random variable by observing the other variables:

$$I(X;Y;Z) = I(X;Y) + I(X;Z|Y)$$
$$= H(X) + H(Y) + H(Z) - H(X,Y) - H(X,Z) - H(Y,Z) + H(X, Y, Z)\,,$$
where $I(X;Y)$ is the mutual information between $X$ and $Y$, while $H$ is the entropy. In other words, co-information~\citep{bell2003coinfo} captures how much information is collectively shared among all variables in a system, by quantifying the extent to which the joint distribution of all variables contributes to our knowledge about individual variables. Even though co-information is a generalization of mutual information, its interpretation differs in that this measure can also be negative in case of $3$ or more variables. As a non-negative generalization of mutual information, total correlation (TC)~\citep{watanabe1960tc} is defined as the difference between the sum of the individual entropies of the variables and their joint entropy:

$$TC(X_1, \dots , X_n) = \sum_{i=1}^n H(X_i) - H(X_1, \dots , X_n).$$

In practice, TC measures the degree of interdependence among a set of variables by comparing their total entropy with the entropy of the joint distribution. In equivalent terms, this corresponds to measuring the amount of information that each individual variable contributes to the collective information represented by the joint distribution. Therefore, TC would be 0 if and only if the variables are independent. On the other hand, TC is maximized when all variables are completely determined by one of them. Similarly, dual total correlation (DTC)~\citep{sun1975dtc} or binding information~\citep{abdallah2012binding} is non-negative, but, instead of being bounded by the sum of the entropies, it is bounded below by 0 and above by the joint entropy:

$$DTC(X_1, \dots , X_n) = H(X_1, \dots , X_n) - \sum_{i=1}^n H(X_i | X_1, \dots , X_{i-1}, X_{i+1}, X_n).$$

Building on these measures, O-information~\citep{rosas2019oinfo} or enigmatic information~\citep{james2011enigm} quantifies higher-order interdependencies by the difference between TC and DTC, which can therefore result in negative scores for synergistic systems:

$$\Omega(X_1, \dots , X_n) = TC(X_1, \dots , X_n) - DTC(X_1, \dots , X_n).$$

While those measures offer valuable insights, a range between 0 (no constraint) and 1 (total dependence) would represent an intuitive scale for a measure of higher-order constraint, helping the interpretability and usability of the measure across different contexts. To this end, DTC can be normalized to be in the range $[0,1]$ by dividing by the joint entropy. Additionally, it might be desirable for a measure of higher-order dependence to reduce to a pairwise measure of dependency in case there are only two entities involved. Among the measures reviewed above, this is true for co-information, which naturally reduces to the mutual information in case it is applied to only two variables.

As for the measure defined in Equation~\ref{eq:general}, it involves the individual entropies as well as the entropy of the joint distribution. In this sense, our measure extends the total correlation by considering also the contribution of lower-order dependencies. Furthermore, our formulation assesses the degree of interdependence among the variables by estimating on average how much information each variable contributes to the collective understanding of the system. Finally, our measure returns 0 in case of no constraint and 1 in case of total dependence.

\newpage
\section{Null Model}
\label{a:nullmodel}

The null model is defined as follows: for each topical issue in the system under examination, the number and sizes of opinion groups remain fixed, but the individual memberships are randomly permuted. This random permutation effectively creates a system with  properties similar to the one under analysis. By averaging the results over multiple iterations of the null model, we obtain the expected alignment scores $\langle A_{\text{null}} \rangle$ for each $k$, and the expected maximum alignment curve for the system. These expected values account for multiway alignment purely attributable to chance.

Consequently, evaluating the multiway alignment scores $A$ (Equation~\ref{eq:general}) obtained from the real-life system involves comparing them to the expected multiway alignment scores $\langle A_{\text{null}} \rangle$ derived from the null model. Specifically, if the multiway alignment scores obtained from the real-life system fall within the $(1-\alpha) \%$ confidence interval of the scores generated by the null model, we can conclude that the alignment observed in the system is not statistically significant at the significance level $\alpha$. Conversely, the greater the difference between the multiway alignment score from the real-life system and the $(1-\frac{\alpha}{2})$-th percentile of the scores obtained from the null model, the stronger the evidence supporting the presence of high alignment in the system.

Therefore, we can adjust the multiway alignment measure defined in Equation~\ref{eq:general} to account for the effect attributed to chance. To achieve this, we follow a similar logic as in \citet{alinormalization} and subtract the expected alignment scores obtained from the null model from the observed multiway alignment scores in the real-life system. This subtraction effectively removes the influence of chance alignment. Subsequently, to ensure that the adjusted measure of alignment falls within the $[0, 1]$ range, we renormalize it by dividing by $1 - \langle A_{\text{null}} \rangle$.

The net multiway alignment score 

\begin{equation}
    A_{\text{net}} = \frac{A - \langle A_{\text{null}} \rangle}{1 - \langle A_{\text{null}} \rangle}
\end{equation}
serves as a robust and interpretable metric for quantifying the alignment of opinion groups across multiple topics, enhancing our ability to discern statistically significant alignment patterns within complex systems. The net alignment score represents, for each $k$, the excess alignment that cannot be attributed to chance, thus eliminating the implicit dependency of the multiway alignment score on the number of topics $k$.

Our empirical results show that $\langle A_{\text{null}} \rangle$ can be far from 0 when utilizig normalized mutual information (NMI), whereas $\langle A_{\text{null}} \rangle$ is consistently close to 0 when using AMI.

\newpage
\section{Multiway Alignment in Finnish Twitter Before Elections}
    \label{a:results-twitter}

\begin{figure}
    \centering
    \includegraphics[width=\linewidth]
    {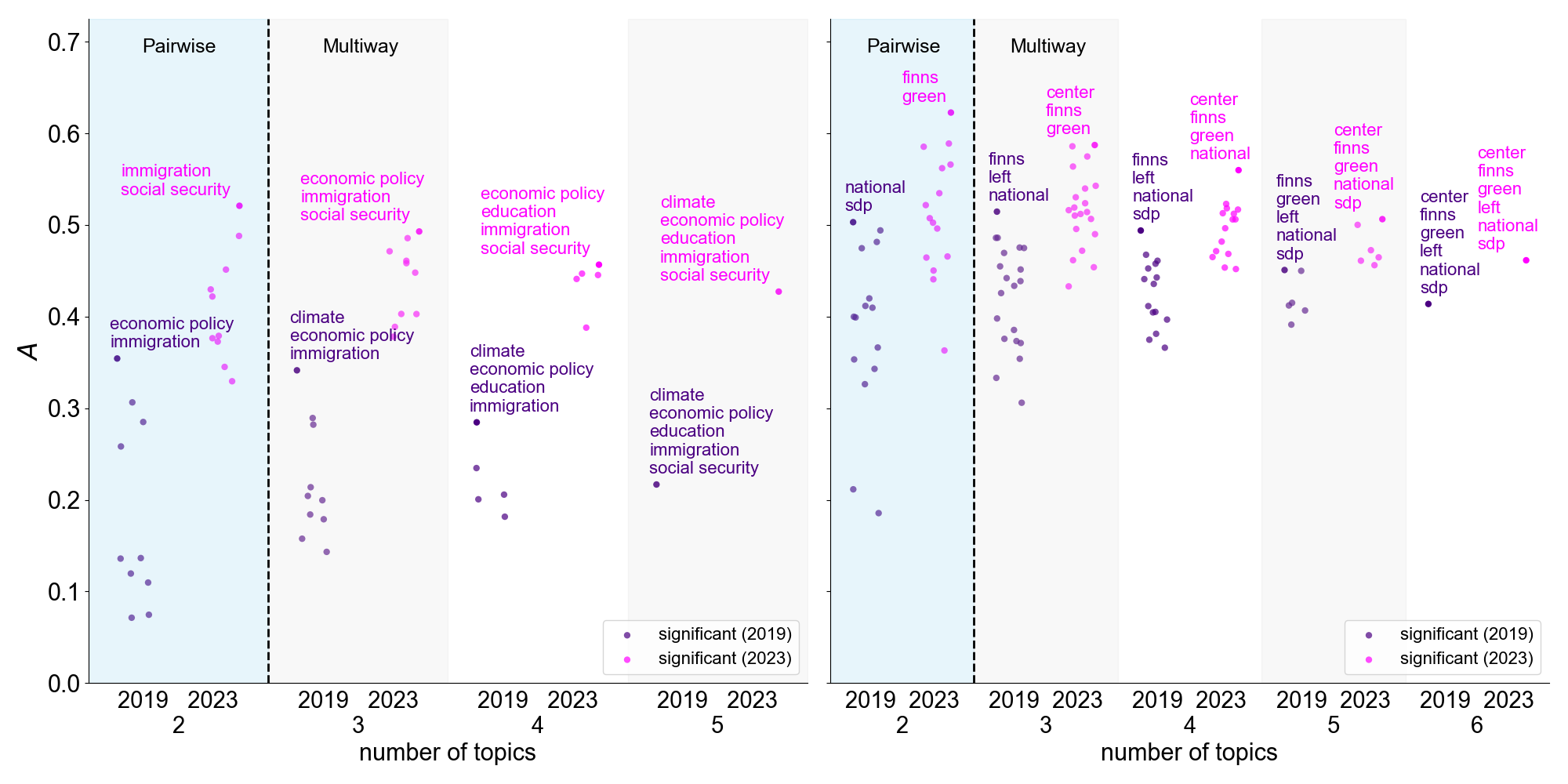}
    \caption{The comparison of multiway alignment in Finnish Twittersphere before 2019 and before 2023 elections shows an overall increase in alignment, as well as changing patterns in the alignment induced by issues and parties discussed before elections.}
    \label{fig:twitter-double-subplots}
\end{figure}

Next, we explore multiway alignment patterns in Twitter discussions in the months leading up to the 2019 and the 2023 Finnish political elections. We show these results in Figure~\ref{fig:twitter-double-subplots}. The spectrum of the multi-alignment across selected salient issues (left panel in Figure~\ref{fig:twitter-double-subplots}) and party preferences for the major five parties (right panel in Figure~\ref{fig:twitter-double-subplots}) unveils evident trends and shifts in inter-topic and inter-party relationships.

Both panels in Figure~\ref{fig:twitter-double-subplots} confirm the previously observed significant rise in pairwise alignment in the Finnish online discourse~\citep{salloum2024anatomy}. However, they also show a concerning increase in higher-order alignment across all $k$, underscoring a heightened interdependency among political elements discussed online. Exemplification of this are the instances of $k=3$ and $k=4$ in Figure~\ref{fig:twitter-double-subplots} (left panel), where the least aligned $k$-tuple of issues in 2023 surpasses the alignment of the most aligned $k$-tuple of issues in 2019, indicating a substantial increase in alignment as measured through political discussions on Twitter.

While an increase in higher-order alignment is also evident when analyzing party positions, the magnitude of this rise is notably smaller compared to the trends observed among issues. The emergence of alignment within discussions around political parties, particularly in a multi-party system where coalition governments are commonplace, is somewhat anticipated. However, the substantial upsurge in higher-order alignment across issues signifies a shift towards a more partisan-centric political landscape. Moreover, the difference in the evolution of alignment over time between issues and party preferences suggests a transformation in the nature of political discourse. The increasing alignment among issues indicates a trend where discussions are increasingly influenced by partisan ideologies rather than a mere representation of varied political party preferences. This shift might imply a growing polarization of political viewpoints across a diverse range of issues, emphasizing the intertwining of diverse concerns into a more ideologically driven political sphere.

Figure~\ref{fig:twitter-double-subplots} also reveals shifts in issue relevance over time. In 2019, the most aligned pair of issues comprised economic policy and immigration, whereas in 2023, immigration and social security emerged as the most aligned pair, with economic policy replacing climate change in the most aligned 3-tuple. The analysis of discussions around parties in 2019 hint at a constraint between the opinion of Twitter users on SDP, the election winner, and the National Coalition Party. In contrast, the online discourse surrounding 2023 elections showcases a correlation between the perspectives on the Finns Party and the Green Party, which occupy opposing roles in the current government. Beyond government-opposition dynamics, this shift may also reflect a broader transition from the traditional left-right political identities that were dominant in 2019 to a nationalist-internationalist divide by 2023.

\newpage
\section{Multiway Alignment of Elites: U.S. House}
    \label{a:results-house}

\begin{figure}
    \centering
    \includegraphics[width=1.1\linewidth, trim={5cm, 0cm, 0cm, 0cm}, clip]
    {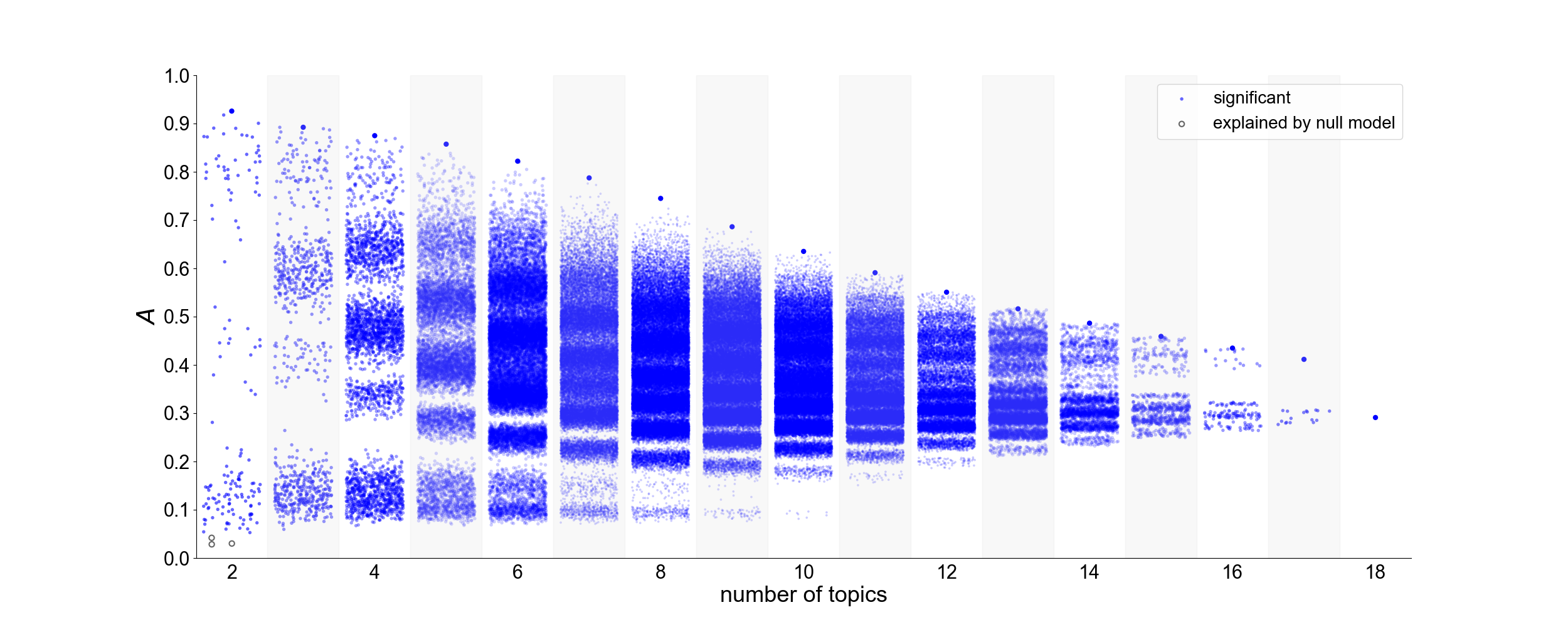}
    \caption{The spectrum of alignment measured from U.S. House in 2014 shows clusters of $k$-tuples that all reach high level of multiway alignment, as well as a value of higher-order alignment at $k=18$ still very far from 0.}
    \label{fig:usa_house_2014}
\end{figure}

Both the House and Senate exhibit clear patterns of alignment that could originate from party lines within these legislative chambers. The alignment spectrum might exhibit clustering patterns as different $k$-tuples reach similar alignment scores. Such clusters are clearly visible, for example, from the spectrum of multiway alignment of U.S. House in 2014 (Figure~\ref{fig:usa_house_2014}) at different values of $k$ ($k=2$, $k=3$, $k=4$, $k=5$, $k=16$). As the most aligned $k$-tuples are not isolated extreme occurrences, the distinguishable clusters of alignment scores suggest that members of the House might consistently align even on a wider set of issues.

\newpage
\section{Multiway Alignment in Finnish Parliament Before and During COVID Pandemic}
    \label{a:results-finland}

\begin{figure}[!t]
    \centering
    \includegraphics[width=\linewidth,trim={0 0 0 0},clip]
    {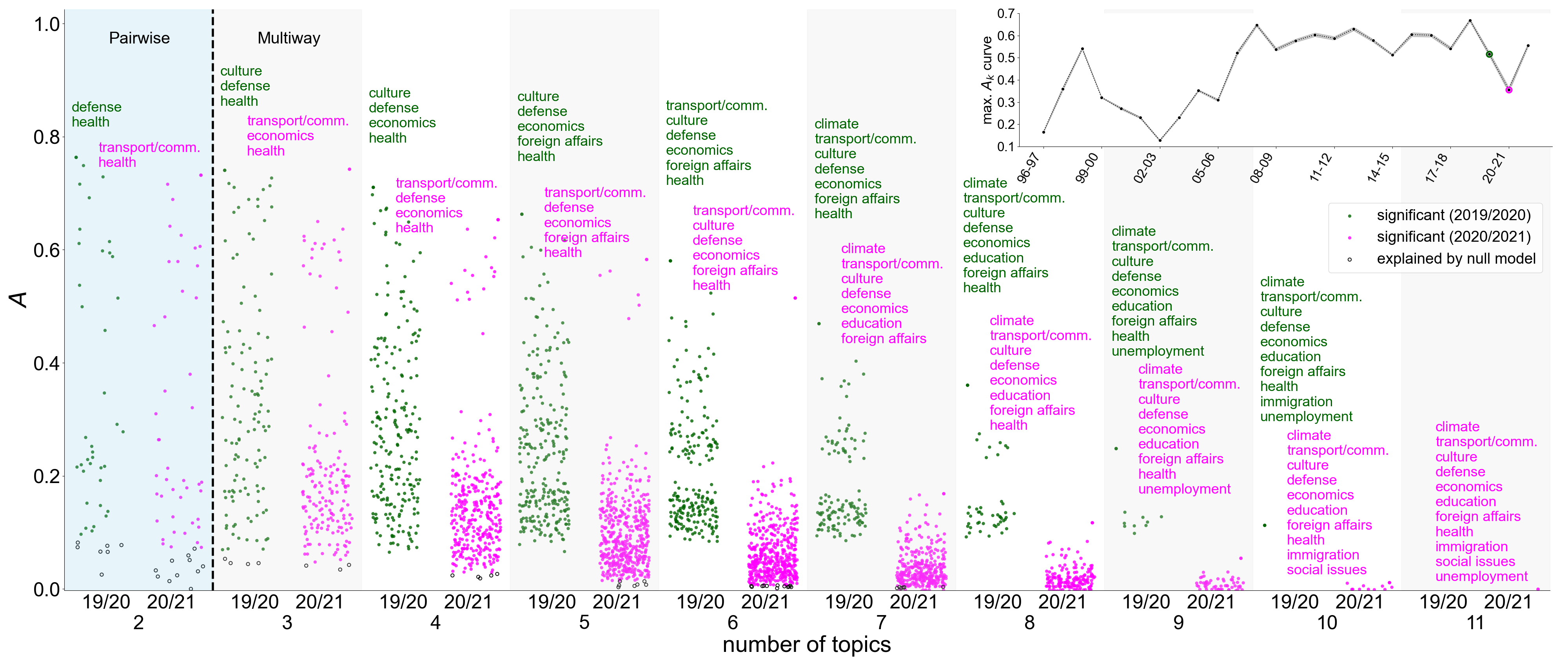}
    \caption{Multiway alignment in Finnish Parliament over time, detailing the patterns before and during the COVID pandemic (2019--2020 and 2020--2021).}
    \label{fig:mainfig-finland}
\end{figure}

The analysis of multiway alignment within the Finnish Parliament offers insights into the shifting political dynamics before and during the COVID-19 pandemic, and reveals a known phenomenon that a purely pairwise analysis would miss. The line chart in Figure~\ref{fig:mainfig-finland} shows the evolution over time of the overall alignment, summarized through the area under the maximal alignment curve. The two points highlighted in the line chart show how the emergence of the pandemic and the subsequent declaration of a state of emergency led to a sharp decrease in alignment in 2020--2021.

A comparison of the multiway alignment spectrum for the 2019--2020 and 2020--2021 parliamentary periods is shown in the main panel in Figure~\ref{fig:mainfig-finland}. Before the onset of the pandemic, the spectrum of the multiway alignment within the Finnish Parliament displayed a funnel shape that we commonly observe in legislative systems, where the alignment across different topics reflects the dynamics of policymaking and political discourse. The funnel shape of multiway alignment indicates a gradual decrease in alignment as the number of topics in the analysis increases. Nevertheless, the values can remain at values above zero also at higher orders (e.g., $k=9$ and $k=10$ in 2019--2020), demonstrating alignment that could originate from adherence to party lines, whereby members from the same party (or coalition) vote in a similar way on a wide range of issues.

The first hint of a reorientation of priorities and focal points within parliamentary deliberations in 2020--2021 is a shift in the most aligned pair of topics, from defense and health in 2019–-2020 to transportation and health in 2020-–2021. However, the manifest shift in the distribution of the scores already at $k=3$ usher in a period of exceptional policy discussions. At $k=3$, the spectrum for 2020--2021 shows two clear clusters of topics, one retaining alignment levels comparable to the previous period, and the other exhibiting markedly low alignment. The same clustering effect becomes increasingly distinct at higher orders ($k=4, 5, 6$): in 2020--2021, six topics -- transportation, culture, defense, economics, foreign affairs, and health -- clearly follow similar alignment patterns as in 2019-2020, while the remaining five topics -- climate, education, immigration, social issues, and unemployment -- become much less aligned, bringing the multiway alignment score at $k=10$ and $k=11$ towards zero.

On one hand, the sudden prominence of topics related to transportation and their alignment with health and economics suggests a focused effort within the Finnish Parliament to navigate the circumstances imposed by the pandemic through initiatives related to free movement restrictions, economic stimulus packages, quarantine and isolation protocols, and wide vaccination efforts~\citep{YLE2020a, YLE2020b}. On the other hand, topics such as climate and immigration could have been perceived as less immediately relevant or lower in priority compared to the urgent challenges posed by the crisis. Overall, the observation of the double-triangle pattern of alignment in the analysis of Finnish Parliament voting supports the claim that, throughout the pandemic, parliamentary discussions diverged from traditional political dynamics, and that the opposition adopted a collaborative approach, refraining from strong criticism of COVID policies proposed by the government~\citep{niemikari2022centralized}. 

\newpage
\section{Use of the \texttt{multiway\_alignment} Python Package}
\label{a:pip-package}

The \texttt{multiway\_alignment} Python package developed in conjunction with this research is available as open-source in GitHub. The code has $90\%$ coverage from unit tests and it is tested on Python $3.10$, $3.11$, and $3.12$. The \texttt{multiway\_alignment} package can be utilized in any Python environment using Python $\geq 3.10$ by following the steps outlined below. The package can be installed directly from the Python Package Index (PyPI) using \texttt{pip} from the command-line interface, by executing the following command:\\

\begin{lstlisting}[style=terminal, caption={Installing from pip. }, label=lst:cmd]
$ pip install multiway_alignment
\end{lstlisting}

Alternatively, the package can be installed by first cloning the repository containing the source code from GitHub and then installing it locally in a chosen directory, with the following commands:\\

\begin{lstlisting}[style=terminal, caption={Installing from source. }, label=lst:cmd_2]
$ git clone https://github.com/letiziaia/multiway-alignment.git
$ cd multiway-alignment
$ pip install .
\end{lstlisting}

Once the package is installed, it can be imported in a Python script or Jupyter environment to perform multiway alignment analysis. The documentation and example usage of the functions in the package is available in GitHub. Below is an example of how to import the package and use it to compute 3-way alignment between topics A, B, and C:\\

\begin{lstlisting}[language=Python, style=mystyle, caption={Example Python Code.}, label=lst:python]
import pandas as pd
# Importing the package
import multiway_alignment.consensus as mac
import multiway_alignment.score as mas

# Load the opinion labels for each issue to a pandas DataFrame
df = pd.DataFrame(
    {
        # on topic A, the first two individuals 
        # have opinion 0, while the following two
        # have opinion 1
        "A": [0, 0, 1, 1],
        "B": [0, 1, 0, 1],
        "C": [1, 0, 1, 0],
    }
)

# get the labels for the consensus partition over topics A, B, and C
partition_labels = mac.get_consensus_labels(opinions=df)

# compute 3-way net alignment score over topics A, B, and C
# using adjusted mutual information without the null model
a = mas.multiway_alignment_score(
    df, which_score="ami", adjusted=False,
)
\end{lstlisting}

\newpage
\section{Alternative Formulation Using Full Consensus Partitions}
\label{s:anmi}

The multiway alignment defined in Equation~\ref{eq:general} compares the opinion partition defined for each topic with the consensus partition defined by the remaining topics. However, given the topics $T_1, \ldots, T_k$, we could have defined multiway alignment utilizing normalized mutual information (NMI) as

\begin{equation}\label{eq:alternative}
    A^* = \frac{1}{k} \sum_{i=1}^{k} \text{NMI}(C(T_1, \ldots, T_k), T_i),
\end{equation}

where NMI is the pairwise normalized mutual information~\citep{nmi}, $C(T_1, \ldots, T_k)$ is the full consensus partition of individuals defined by all the topics $T_1, \ldots, T_k$, and each $T_i$ defines the opinion groups for the $i$-th individual topic.

Mutual information~\citep{nmi} is defined in terms of entropy $H$ as
$$\text{I(X, Y)} = H(X) + H(Y) - H(X, Y),$$
which is symmetric with respect to $X$ and $Y$, and equivalent to
$$\text{I(X, Y)} = H(X) - H(X | Y).$$

By definition, any full consensus partition C perfectly satisfies homogeneity~\citep{homogeneity} with respect to each individual topic $T_i$, since each of the consensus groups only contains members of the same opinion group on the given topic. In terms of conditional entropy, this means that $H(T_i|\text{C}) = 0$ $\forall i$. Therefore, for each topic, the contribution of the mutual information in Equation~\ref{eq:alternative} simplifies to
$$\text{I}(\text{C}, T_i) = H(T_i),$$
which can be normalized with the arithmetic average of the entropies: 

$$
\text{NMI}(\text{C}, T_i) = \frac{2 H(T_i)}{H(T_i) + H(\text{C})} = \frac{2}{1 + \frac{H(\text{C})}{H(T_i)}}.
$$

Since by definition of consensus partition, $H(\text{C}) = H(T_1, \dots , T_k)$,

$$
\text{NMI}(\text{C}, T_i) = \frac{2 H(T_i)}{H(T_i) + H(\text{C})} = \frac{2}{1 + \frac{H(T_1, \dots , T_k)}{H(T_i)}},
$$
which turns Equation~\ref{eq:alternative} into:

$$
A^* = \frac{1}{k} \sum_{i=1}^{k} \frac{2}{1 + \frac{H(T_1, \dots , T_k)}{H(T_i)}}.
$$

This equivalent formulation highlights the interpretation of $A^*$ as a normalized comparison between the joint entropy of the variables and the individual entropies. Specifically, the term $\frac{H(T_1, \dots , T_k)}{H(T_i)} \geq 1$ for any $T_i$. In the extreme case of perfectly overlapping opinion distributions across all the $T_i$, the joint entropy $H(T_1, \dots , T_k)$ equals the entropy of each one of the $T_i$, thus reducing the term $\frac{H(T_1, \dots , T_k)}{H(T_i)}$ to $1$, which results in $A^* = 1$.

Another interpretation of the normalized mutual information with respect to the consensus partition derives from the homogeneity property. By the definition of mutual information, it follows that

$$H(T_i) + H(\text{C}) = H(T_i) + H(T_i, \text{C})$$

and therefore, by the definition of conditional entropy, 

$$H(T_i) + H(\text{C}) = 2 H(T_i) + H(\text{C} | T_i),$$

the formula above can be also expressed in terms of the conditional entropy as

$$\text{NMI}(\text{C}, T_i) = \frac{1}{1 + \frac{1}{2} \frac{H(\text{C} | T_i)}{H(T_i)}}.$$

In particular, $H(\text{C} | T_i) = 0$ (completeness property~\citep{homogeneity}) if and only if the consensus partition does not further disgregate the opinion groups formed at the topic $T_i$. In practice, this means that all of the individuals that belong to the same opinion group on topic $T_i$ are also found in the same consensus group defined by the set of topics. Since the multiway metric we defined in Equation~\ref{eq:alternative} is the arithmetic average of $\text{NMI}(T_i, \text{C})$ over a set of topics, this proves that $A^* = 1$ if and only if the $k$ topics define the exact same partitioning of the individuals.

Finally, variation of information~\citep{variation-info} is defined in terms of entropy $H$ as

$$\text{VI}(X, Y) =  H(X|Y) + H(Y|X),$$
which, by the homogeneity property, simplifies to

$$\text{VI}(T_i, \text{C}) = H(\text{C} | T_i),$$
showing that Equation~\ref{eq:alternative} can also be expressed in terms of the variation of information by using terms of the form

$$\text{NMI}(\text{C}, T_i) = \frac{1}{1 + \frac{1}{2} \frac{\text{VI}(T_i, \text{C})}{H(T_i)}}.$$

This allows to bound the terms, since variation of information is non-negative and equal to 0 when the two partitions $T_i$ and $C$ being compared are identical, which leads to a contribution $\text{NMI}(\text{C}, T_i) = 1$. On the other hand, $\text{VI}(T_i, \text{C})$ is upper-bounded in terms of the number of opinion groups $|C|$, or in terms of the size $N$ of the population in case $|C| = N$, which happens if the consensus partition reduces to a collection of singletons. Therefore

$$0 \leq \frac{\text{VI}(T_i, \text{C})}{H(T_i)} \leq \frac{\log(|C|)}{H(T_i)}$$
implies that

$$0 \leq \text{NMI}(\text{C}, T_i) \leq \frac{{H(T_i)}}{{H(T_i)} + \frac{1}{2} \log(|C|)} \leq 1.$$

\end{document}